\documentclass[paper, 12pt, letterpaper, epsf]{JHEP} 
\usepackage{graphics} 

\def\Bbb{\bf}

\def\C{{\Bbb C}} \def\R{{\Bbb R}} \def\Z{{\Bbb Z}} 
  \def\j{{\Bbb j}}

\def\id{\protect{{1 \kern-.28em {\rm l}}}}

\newcommand{\be}{\begin{equation}} \newcommand{\ee}{\end{equation}}
\newcommand{\bea}{\begin{eqnarray}} \newcommand{\eea}{\end{eqnarray}}
\newcommand{\beann}{\begin{eqnarray*}} \newcommand{\eeann}{\end{eqnarray*}}
\newcommand{\bfig}{\begin{figure}} \newcommand{\efig}{\end{figure}}
\newcommand{\nn}{\nonumber}
\newcommand{\ba}{\begin{array}}\newcommand{\ea}{\end{array}}

\newtheorem{Proposition}{Proposition}[section]

\newtheorem{Theorem}{Theorem}[section] \newtheorem{Lemma}{Lemma}[section]
\newtheorem{Corrolary}{Corrolary}[section]

\newcommand{\bp}{\begin{Proposition}} \newcommand{\ep}{\end{Proposition}}
\newcommand{\bt}{\begin{Theorem}} \newcommand{\et}{\end{Theorem}}
\newcommand{\bl}{\begin{Lemma}} \newcommand{\el}{\end{Lemma}}
\newcommand{\bc}{\begin{Corrolary}} \newcommand{\ec}{\end{Corrolary}}

\title{Holomorphic matrix models}

\author{C. I. Lazaroiu\\ 
Insitut f\"ur Physik\\
Humboldt Universit\"at zu Berlin\\ 
Invalidenstrasse 110, Berlin\\ Germany\\
calin@physik.hu-berlin.de}

\abstract{This is a study of holomorphic matrix models, the matrix models 
which underlie the conjecture of Dijkgraaf and Vafa.
I first give a systematic description of the holomorphic one-matrix model.
After discussing its convergence sectors, I show that certain puzzles related
to its perturbative expansion admit a simple resolution in the holomorphic
set-up. Constructing a `complex' microcanonical ensemble,
I check that the basic requirements of the conjecture (in
particular, the special geometry relations involving chemical potentials) 
hold in the absence of the hermiticity constraint. 
I also show that planar solutions of the holomorphic model 
probe the entire moduli space of the associated algebraic curve. 
Finally, I give a brief discussion of holomorphic $ADE$
models, focusing on the example of the $A_2$ quiver, for which I
extract explicitly the relevant Riemann surface.
In this case, use of the holomorphic model is crucial,  
since the Hermitian approach and its attending regularization
would lead to a singular algebraic curve, thus contradicting the 
requirements of the 
conjecture. In particular, I show how an appropriate regularization of 
the holomorphic $A_2$ model produces the desired smooth Riemann 
surface in the limit 
when the regulator is removed, and that this limit can be described as 
a statistical ensemble of `reduced' holomorphic models. }

\preprint{HU-EP 03/10}

\begin{document}

\tableofcontents


\section*{Introduction}

In the seminal paper \cite{DV1}, Dijkgraaf and Vafa proposed a beautiful
conjecture relating matrix models to closed string theory on
certain noncompact Calabi-Yau spaces. In its strongest form, this is meant as
a relation between the partition function of a certain 
matrix model
and the partition function of Kodaira-Spencer theory 
\cite{BCOV} on a dual
noncompact Calabi-Yau threefold. The large $N$ limit of this
relation also leads to a matrix model description of 
corrections to the Veneziano-Yankielowicz potential of 
certain $N=1$ supersymmetric field theories \cite{DV1, DV2} (moreover, 
the VY potential itself can be recovered from the non-perturbative part of the 
large $N$ matrix integral). In fact, there are now  independent
derivations of the planar version of this conjecture by purely field-theoretic
methods \cite{D_ft,Seiberg, Witten,Gorski}, 
as well as first tests of it beyond the
large $N$ limit \cite{Klemm, DSM}. More recent work on the subject can be
found in \cite{Klemm2, Corrado, Seiberg, Whitham2, Whitham3, Witten2, 
DV4, DV5, Tatar}.

As already pointed out in  \cite{DV1}, a proper formulation of the 
conjecture should be given in terms of `holomorphic matrix models', 
namely some version of matrix models involving `contour' integrals in 
a space of complex matrices. This is very natural 
if one remembers that the conjecture  
was originally derived by considering the 
worldvolume theory of certain topological B-type branes, which is described by
a reduction of holomorphic Chern-Simons theory \cite{Witten_CS}.  
Since the latter is formulated without reference to a metric, 
the proper description of the resulting matrix model 
should not involve a hermiticity constraint. 
This expectation is borne out 
by the fact that the conjecture describes relations in the chiral ring
of the dual field theory \cite{Witten}, and the latter is constrained by 
holomorphy.

In \cite{DV1}, the authors chose the pragmatic approach 
of formulating the conjecture in terms of Hermitian matrix models, 
while pointing out that an appropriate holomorphic formulation 
should be given. The purpose of the present work is to initiate a systematic 
study of such holomorphic models, and show how they naturally 
tie up some loose ends encountered in various investigations and extensions 
of the conjecture.

If naively taken at face value, the 
Hermitian formulation leads to various
problems, some of which were already encountered (and overcome by 
a pragmatic prescription) in work aimed at testing it \cite{Klemm}. 
Among such issues, one can list the following:

(a) Hermitian one-matrix models with generic polynomial potentials of odd
degree (and complex coefficients) are ill-defined, since the real part of such
potentials is not bounded from below along the real axis. However, such models
are naturally required by the conjecture, which is supposed to apply without
constraints on the degree of the potential. This issue was encountered in the
work of \cite{Klemm}, while performing a one-loop test of the
conjecture for a {\em cubic} potential. As shown in Section 2 and Appendix 2, 
the pragmatic approach followed in \cite{Klemm} admits a 
natural justification in the holomorphic setup.

(b) The large $N$ spectral density $\omega_0(z)=tr\frac{1}{z-M}$ of a
    Hermitian one-matrix model (with matrix $M$) can only have cuts along 
    the real
    axis. This means that the cuts of the hyperelliptic Riemann surface of
    \cite{DV1} would be constrained to lie on the real line. In the Hermitian
    formulation, the conjecture would then imply that the dual
    Calabi-Yau space is constrained in a similar manner. Moreover, the
    Hermitian formulation leads to numerous problems in matching parameters
    and moduli spaces, since
    a polynomial of degree $n$ always has $n$ complex roots, 
    but need not have $n$ {\em real} roots. As clear from the work of 
    \cite{DV1} (and occasionally pointed out in subsequent work by the same 
authors), what is needed is a holomorphic extension of the Hermitian 
matrix model which would allow for an enlarged family of planar limits ---
    namely, such a model should produce solutions whose large $N$ 
    eigenvalue distributions can be supported along arbitrary
    one-dimensional curve segments in the complex plane. 

In the present note, I show how these and related issues are solved by a 
direct analysis of holomorphic matrix models, whose construction follows 
a suggestion already made in \cite{DV1}. 
In Sections 1 and 2, I explicitate the
non-perturbative definition of such models
\footnote{I should state from the outset that holomorphic matrix models
are not the same as 
the so-called `complex matrix models'  \cite{complex} 
sometimes encountered in the
matrix model literature.
Rather, they are a sort of `square roots' of such  
models.} 
and study their convergence sectors, thereby refining one side of the 
conjecture. I also show (in Appendix 2) that 
issues like those encountered in
\cite{Klemm} admit a natural resolution in this framework (which {\em does} 
recover the prescription used in that paper). 
Section 3 extracts the loop equations, equations of motion and the planar
limit of such models, showing that most of the standard analysis 
employed for the Hermitian model carries through with certain modifications.
Consideration of the large $N$ limit leads to the algebraic curve of
\cite{DV1}, which is now {\em unconstrained} by any conditions on the location 
of the cuts. In fact, one can give a `reconstruction 
theorem', which ensures that the 
holomorphic model probes the entire moduli space of this algebraic curve. 
This shows explicitly 
how such models solve the second issue 
listed above. Up to some details, the reconstruction result boils down 
to the well-known relation between the Riemann problem and singular 
integral equations. From this perspective, holomorphic matrix models give a 
sort of `quantization' of the classical Riemann problem. 

The validity of the conjecture rests crucially on the special geometry
relations mentioned in \cite{DV1}, a more detailed justification of which was 
latter given in \cite{Whitham} (though only for the Hermitian case). 
To give a clear proof of such relations at the holomorphic level, 
Section 4 constructs a `complex' microcanonical ensemble by 
introducing chemical
potentials and performing a Legendre transform, thereby obtaining a generating
functional which depends on {\em averaged} filling fractions; this allows 
one to show that the special geometry relations of \cite{DV1} are a direct
consequence of standard equations expressing chemical potentials 
in terms of the microcanonical generating function. In particular, these 
relations give the finite $N$ analogue of the special geometry 
constraints.

Section 5 considers holomorphic $ADE$ models, focusing on a detailed 
analysis of the $A_2$ model. In this case, the problems one encounters by 
working naively with a Hermitian constraint are considerably more severe 
than in the one-matrix case. In fact, the Hermitian approach of 
\cite{Kostov_ADE} (with its attending regularization) 
would lead to a singular curve, which would always be constrained to have 
a certain number of double points.
This would completely miss 
some of the gauge theory physics extracted in \cite{Cachazo, Radu}
More precisely, such a constraint would force certain filling fractions 
to be identically zero, thereby contradicting the existence of  
the gaugino condensates obtained in \cite{Cachazo}. As shown in Section 5, 
this issue is resolved quite naturally in the {\em holomorphic} $A_2$
model, by considering a regularization which is natural in that set-up. 
When removing the regulator, one obtains an ensemble 
of reduced holomorphic models, whose planar limit allows for non-vanishing
values of {\em all} filling fractions. It is the large $N$ limit of this 
ensemble which is relevant for the conjecture of \cite{DV1}.

\section{Construction of holomorphic one-matrix models 
and their eigenvalue representation} 

Following a suggestion made in \cite{DV1}, we start by constructing 
holomorphic one-matrix models as multidimensional `contour' integrals 
in a space of complex matrices. The construction is quite similar 
to that of a Hermitian model, except that the hermiticity constraint is 
replaced by a more general condition on the eigenvalues. After defining the 
model and studying its gauge-invariance, we extract an eigenvalue
representation by choosing an appropriate multidimensional `contour'. 
With this choice, one obtains an integral expression for the partition 
function which is formally identical to that of a Hermitian model, except that 
integration is performed over eigenvalues lying along an open path 
in the complex plane. 

\subsection{The partition function}

Fixing a positive integer $N$, we let  $Mat_N(\C)$ denote 
the set of all $N\times N$ complex matrices. For any such matrix $M$, we let 
$p_M(\lambda)=det(\lambda I-M)$ denote its characteristic polynomial.
Define a subset ${\cal M}$ of $Mat_N(\C)$ as follows:
\be
{\cal M}:=\{M \in Mat_N(\C)~|~p_M(\lambda)~~{\rm has~distinct~roots}~\}~~.
\ee
This is an open submanifold of $Mat_N(\C)$,
consisting of matrices which 
are diagonalizable by general linear transformations:
for every $M$ in ${\cal M}$, there exists an $S\in GL(N,\C)$ such that: 
\be
\label{diag}
SMS^{-1}=D:={\rm diag}(\lambda_1\dots \lambda_N)~~,
\ee
where $\lambda_j$ are the roots of $p_M(\lambda)$. The later coincide with the 
eigenvalues of $M$, so that its spectrum is given by:
\be
\sigma(M)=\{\lambda_1\dots \lambda_N\}~~.
\ee

Consider a connected, noncompact and boundary-less submanifold $\Gamma$
of ${\cal M}$ of real dimension equal to the 
complex dimension of ${\cal M}$:
\be
\label{dimGamma}
dim_\R \Gamma=dim_\C{\cal M}=dim_\C Mat_N(\C)=N^2~~.
\ee 
Also consider the standard holomorphic symplectic form on $Mat_N(\C)$:
\be
\label{w_def}
w=\wedge_{i,j} dM_{ij}~~.
\ee
The sign of the right hand side of course depends on the ordering of pairs 
$(i,j)\in \{1\dots N\}\times \{1\dots N\}$, and we shall use the lexicographic 
ordering: 
\be
w=dM_{11}\wedge \dots \wedge 
dM_{1N}\wedge dM_{21}\wedge \dots \wedge dM_{2N} \wedge \dots 
\wedge dM_{N1}\wedge \dots \wedge dM_{NN}~~.
\ee
This is implicit in all such expressions encountered below. 

Fixing a polynomial $W(z)=\sum_{m=0}^{n}{t_m z^m}$ with complex coefficients, 
we define the {\em holomorphic one-matrix model} by the partition function:
\be
\label{hmm}
{\tilde {\cal Z}}_N(\Gamma,t)=\frac{1}{{\cal N}}
\int_{\Gamma}{w e^{-N tr W(M)}}~~,
\ee
where ${\cal N}$ is a normalization factor to be fixed below. 

\subsection{Gauge invariance}

Consider the $GL(N,\C)$ action $\tau$ on $Mat_N(\C)$ 
given by similarity transformations:
\be
\label{sim}
\tau(S)M:=SMS^{-1}~~,~~S\in GL(N,\C)~~.
\ee
This clearly stabilizes ${\cal M}$; in particular, note that 
the characteristic polynomial of $M$ is $\tau$-invariant:
\be
p_{\tau(S)M}(\lambda)=p_M(\lambda)~~.
\ee

The action ${\cal S}(M)=N tr W(M)$ 
of our model is obviously invariant with respect to
(\ref{sim}). The same is true of the holomorphic measure $w$:
since $[\tau(S)M]_{ij}=a_{ij,kl} M_{kl}$, with 
$a_{ij,kl}=S_{ik}(S^{-1})^T_{jl}$ 
(i.e. $a=S\otimes (S^{-1})^T$), we have $det(a)=1$ and:
\be
\label{w_invar}
\tau(S)^*(w)=w~~. 
\ee 
It follows that the model admits the $GL(N,\C)$ gauge-invariance (\ref{sim}),
provided that one chooses the integration manifold $\Gamma$ such that it is 
stabilized by the gauge-group action.  

{\bf Observation}
Choosing $S$ to be a permutation matrix (i.e.
$S_{ij}=\delta_{j\sigma(i)}$ with $\sigma$ a permutation on $N$ elements), 
equation (\ref{w_invar}) shows that the holomorphic 
measure (\ref{w_def}) is invariant under joint 
permutations of indices:
{\footnotesize \bea
&& dM_{11}\wedge \dots \wedge 
dM_{1N}\wedge dM_{21}\wedge \dots \wedge dM_{2N} \wedge \dots 
\wedge dM_{N1}\wedge \dots \wedge dM_{NN}=\\
&& dM_{\sigma(1)\sigma(1)}\wedge \dots \wedge 
dM_{\sigma(1)\sigma(N)}\wedge dM_{\sigma(2)\sigma(1)}
\wedge \dots \wedge dM_{\sigma(2)\sigma(N)} \wedge \dots 
\wedge dM_{\sigma(N)\sigma(1)}\wedge \dots \wedge dM_{\sigma(N)\sigma(N)}~~,\nn
\eea} 
a relation which can also be checked directly.

\subsection{Eigenvalue representation}

One can derive an eigenvalue representation of (\ref{hmm}) as follows. 
Let $\gamma:\R\rightarrow \C$ be an open curve in the complex plane, 
which we assume to be immersed and without self-intersections. 
We shall take $\Gamma$ to be  the following subset of ${\cal M}$:
\be
\Gamma(\gamma):=\{M\in {\cal M}~|~\sigma(M)\subset \gamma\}~,
\ee
which obviously satisfies (\ref{dimGamma}). We denote 
${\tilde {\cal Z}}_N(\Gamma(\gamma),t)$ by ${\tilde {\cal Z}}_N(\gamma,t)$. 
Then the integral over gauge-group orbits can be performed as 
explained in Appendix 1, with the result:

\be
\label{ev_vol}
{\tilde {\cal Z}}_N(\gamma,t)=\frac{1}{{\cal N}}(-1)^{N^2(N-1)/2}
\frac{1}{N !}~hvol(H) {\cal Z}_N(\gamma,t)~~,
\ee 
where $hvol(H)$ is the 
`holomorphic volume' of the complex homogeneous space 
$H=GL(N,\C)/(\C^*)^N$ (see Appendix 1) and 
\be
\label{ev}
{\cal Z}_N(\gamma,t)=
\int_\gamma d\lambda_1 \dots \int_\gamma d\lambda_N
\prod_{i\neq j}{(\lambda_i-\lambda_j)} e^{-N\sum_{j=1}^N{W(\lambda_j)}}~~.
\ee
is the {\em eigenvalue representation} of our model. 
The holomorphic volume $hvol(H)$ will be discarded 
(together with the sign prefactors) by choosing 
${\cal N}=(-1)^{N^2(N-1)/2}
\frac{1}{N !}~hvol(H)$ in (\ref{hmm}).
Expression (\ref{ev}) is formally identical with the eigenvalue 
representation of the
Hermitian matrix model, except that the eigenvalue integral is performed 
along the contour $\gamma$ in the complex plane. The pragmatic reader 
can take (\ref{ev}) as the definition of our model.

\section{Convergence sectors}

It is clear from expression (\ref{ev}) that convergence of our partition 
function depends on the choice of  $\gamma$. In this section, we shall 
characterize the `good' choices of $\gamma$ in terms of certain asymptotic
sectors of the model, described in terms of cones partitioning the complex
plane. Such cones can be identified by performing an 
elementary analysis of the behavior of the integrand. As we shall see 
below, this allows us to make non-perturbative sense of models with 
polynomial potentials of odd degree.

To extract the relevant behavior, let $z=re^{i\theta}$ with 
$r>0$ and $\theta\in \R/2\pi \Z$ and $t_j=r_je^{-i\theta_j}$ with $r_j>0$ 
and $\theta_j\in \R/2\pi \Z$. Then the potential takes the form:
\be
\label{W_r}
W(z)=t_0+\sum_{j=1}^n{r_jr^j\cos(j\theta-\theta_j)}+i 
\sum_{j=1}^\infty{r_jr^j\sin(j\theta-\theta_j)}~~,
\ee
and the behavior of $|e^{-N W(z)}|=e^{-N ReW(z)}$ for 
$r\rightarrow \infty$ is dictated by the 
contribution $r_{j_0}r^{j_0}\cos(j_0\theta-\theta_{j_0})$, where 
$j_0=j_0(\theta)$ is the
largest $j$ such that  $r_j\cos(j\theta-\theta_{j})\neq 0$. Namely, 
$e^{-N W(z)}$ is exponentially decreasing/increasing depending on whether 
$\cos(j_0\theta-\theta_{j_0})$ is positive/negative. 
In the very non-generic case 
when $r_j\cos(j\theta-\theta_{j})$ vanishes for all $j=1\dots n$, the quantity 
$e^{-N W(z)}$ oscillates indefinitely as $r\rightarrow \infty$.  

Let us fix $\theta$ such  that $\cos(n \theta-\theta_n)\neq 0$. 
Then $e^{- N W(z)}$ is exponentially
decreasing as $r\rightarrow \infty$ if and only if
$\cos(n\theta-\theta_n)>0$,
which gives:
\be
\theta=\frac{\alpha+\theta_n}{n}+\pi \frac{2 k}{n}~~{\rm with}~~k= 0\dots
n-1 ~~{\rm and}~~\alpha\in (-\pi/2,\pi/2)~~.
\ee
This defines $n$ angular sectors (=open cones with apex at the origin) 
in the complex plane, which we denote by 
$A_k$, $k=0\dots n-1$. We also define complementary sectors $B_k$ through:
\be
\theta=\frac{\alpha+\theta_n}{n}+\pi \frac{2 k+1}{n}~~{\rm with}~~k= 0\dots
n-1 ~~{\rm and}~~\alpha\in (-\pi/2,\pi/2)~~;
\ee
these are the sectors for which $\cos(n \theta-\theta_n)<0$.
The sectors 
$A_k$ and $B_k$ arise consecutively with respect to the trigonometric
order and are separated by rays originating at $z=0$.

\begin{figure}[hbtp]
\begin{center}
\scalebox{0.5}{\input{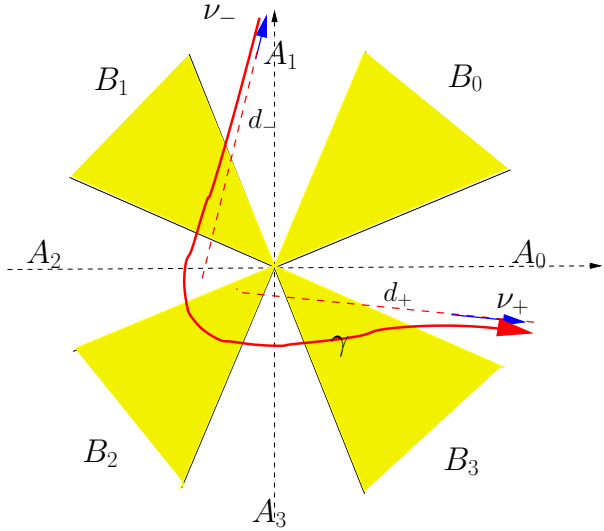}}
\end{center}
\caption{Convergence sectors of the holomorphic matrix model. We show the 
case $n=deg W=4$, with $\theta_4=0$ and a contour belonging to the 
sector $(k_-,k_+)=(1,0)$.}
\label{sectors}
\end{figure}

To make the integral (\ref{ev})
absolutely convergent, we shall require that 
the curve $\gamma$ asymptotes to some straight lines 
$d_\pm(t)=\pm \nu_{\pm}t+\mu_\pm$ for $t\rightarrow \pm \infty$, 
such that the corresponding asymptotes lie inside two of the convergence
sectors $A_k$. That is, we require: 
\bea
\label{as}
&\exists& \lim_{t\rightarrow \pm \infty}{
(\gamma(t)\mp \nu_{\pm}t-\mu_\pm})=0~~\\
&\exists& \lim_{t\rightarrow \pm \infty}{\gamma'(t)}:=\pm \nu_\pm~~.\nn
\eea
for some $\nu_{\pm}\in A_{k_\pm}$ and some $\mu_\pm \in \C$. 
Here $\gamma'(t)=\frac{d\gamma(t)}{dt}$ and 
$k_{\pm }\in \{0\dots n-1\}$. Since the integrand of (\ref{ev}) is
holomorphic, it immediately 
follows that the partition function is independent
of the choice of $\gamma$ as long as this contour 
has asymptotic behavior (\ref{as}) with $\nu_\pm$ belonging to fixed 
sectors $A_{k_\pm}$. In particular, the integral does not 
depend on the precise values of $\nu_\pm$ and $\mu_\pm$, but only on the 
convergence sectors $A_{k_+}$ and $A_{k_-}$.
Therefore, we obtain $n^2$ possible values ${\cal Z}(k_-,k_+,t)$, 
indexed by the sectors $A_{k_-}$ and $A_{k_+}$ (figure \ref{sectors}). 
A complete definition of the model requires that we specify one of the 
`phases' $(k_-,k_+)$, together with the potential $W$. 

It is also clear that ${\cal Z}(k_-,k_+,t)$ vanishes if $k_+=k_-$, and 
that we have the relation:
\be
{\cal Z}(k_-,k_+,t)=(-1)^N{\cal Z}(k_+,k_-,t)~~,
\ee 
which results upon reversing the orientation of $\gamma$. Therefore, it 
suffices to consider the $n(n-1)/2$ `phases' indexed by pairs $(k_-,k_+)$ 
with $k_->k_+$.

\subsection{Scaling symmetry} 

Let $q$ be a non-vanishing complex number. 
Writing $z:=\frac{x}{q}$, we have:
\be
W(z)=W_q(x)~~,
\ee
where $W_q(x)=\sum_{j=0}^n{t^{(q)}_j x^j}$, with 
$t^{(q)}_j:=\frac{t_j}{q^j}$.
Performing the change of variables $\lambda:=\frac{\mu}{q}$ 
in (\ref{ev}) gives:
\be
\label{ss}
{\cal Z}_N(\gamma,t)=\frac{1}{q^{N^2}}{\cal Z}_N(\gamma_q,t^{(q)})~~,
\ee
where $\gamma_q$ is the path defined through:
\be
\gamma_q(t)=q \gamma(t)~~
\ee
for all real $t$.

For the choice $q_n:=r_n^{1/n}e^{-\frac{i\theta_n}{n}}$, 
the change of variable $z:=\frac{x}{q_n}$ gives $t^{(q_n)}_n=1$,
so the transformed potential $W_{q_n}$ has unit leading coefficient. 
Hence the model depends `trivially' on the parameter $t_n$, and we 
can set $t_n=1$ and $\theta_n=0$ without loss of generality. 

Also note that choosing $q:=\alpha^k$ with $\alpha=e^{\frac{2\pi i}{n}}$ 
and $k$ an integer allows one to rotate 
$\gamma$ by any multiple of $\frac{2\pi }{n}$. Since this does not change the 
convergence sectors (because $t_n^{(\alpha^k)}=t_n$), it leads to the relation:
\be
{\cal Z}_N(k+k_- ,k+k_+,\{t_m\})=\alpha^{N^2 k}
{\cal Z}_N(k_- ,k_+ ,\{\alpha^{mk}t_m\})~~.
\ee
Hence it suffices to consider the `phases' $(k_-,0)$ with 
$k_-=1\dots n-1$.

\begin{figure}[hbtp]
\begin{center}
\scalebox{0.4}{\input{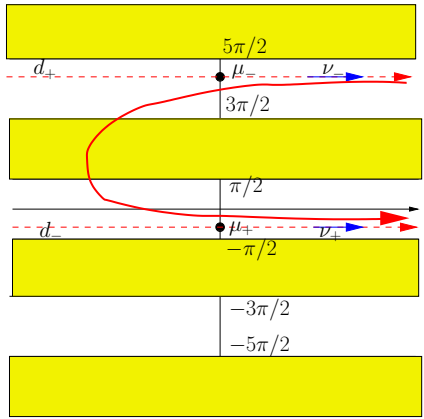}}
\end{center}
\caption{Convergence sectors for an entire potential $W=e^z$, 
and a good choice of contour for such a model. The filled-in regions are
forbidden sectors for $\mu_\pm$.}
\label{entire}
\end{figure}

{\bf Observation:}  
If one increases the degree of $W$, then the 
convergence sectors become progressively narrower. 
Allowing for the case $n=\infty$ (i.e. for an entire function $W$) 
is often a useful device in the theory of matrix models
(the best known example is the matrix model of Kontsevich \cite{Kontsevich}
and its generalizations). In this case, the convergence 
structure of the holomorphic model can be quite different from that 
discussed above, and depends on the precise asymptotic behavior of the entire 
function $W$. A simple example is provided by the choice $W(z)=e^z$, which
gives $|e^{-N W(z)}|=e^{-N Re~W(z)}=e^{-N e^x\cos(y)}$, where $z=x+iy$ with 
$x,y$ real. Then the convergence sectors are horizontal strips 
defined by the condition:
\be
\cos(y)>0\Longleftrightarrow y\in (-\frac{\pi}{2}+2\pi k, \frac{\pi}{2}
+2\pi k)~~,~~k\in \Z~~. 
\ee 
Correspondingly, we obtain a convergent partition function by taking $\gamma$ 
to satisfy (\ref{as}), where now $\nu_\pm>0$ and $\mu_\pm$ 
belong to two such bands (figure \ref{entire}).  This example should serve 
as warning against the idea that one can recover models with power 
series potentials as naive limits of polynomial models.

\subsection{Even and odd degree potentials, and the Hermitian matrix model}

\subsubsection{The case $n=$even}

In the case $n=$even, the convergence sectors of the model appear in pairs 
$A_k$, $A_{k+n/2}$ which are symmetric with respect to the inversion
$z\rightarrow -z$. By the discussion above, we can 
take $t_n$ positive without loss of generality. 
Then $\theta_n=0$ and the two sectors $A_0$ and $A_{n/2}$ lies opposite to
each other and are bisected by the positive real axis.  
Picking $k_-=n/2$ and $k_+=0$, we 
can describe the `phase' ${\cal Z}(n/2,0,t)$ by taking the curve $\gamma$ 
to coincide with the real line (with its usual orientation). Then the partition
function (\ref{ev}) reduces to the usual eigenvalue representation of the 
Hermitian one-matrix model. This gives a partial 
justification for the formulation used in \cite{DV1}. 

\subsubsection{The case $n=$odd}

In this case, one has an odd number of convergence sectors $A_k$, whose
images under the reflection $z\rightarrow -z$ are the `bad' sectors 
$B_{k+[n/2]}$. 
Picking $t_n$ to be positive as above, it is clear that the matrix model
{\em cannot} be defined by choosing $\gamma=\R$, since the associated 
integral would diverge; this amounts to the basic observation that the 
Hermitian matrix model is ill-defined for generic 
complex polynomial potentials of odd degree.
However, one can certainly make sense of the holomorphic model 
by working with any of the good `phases' $(k_-,k_+)$ --- 
one simply picks a contour $\gamma$ such that $\nu_+$ and $\nu_-$ 
belong to some of the convergence sectors $A_k$. 

The fact that the Hermitian model cannot be relevant 
for odd degree potentials is related to problems observed
in \cite{Klemm}, which in that paper were avoided by declaring
that certain matrices occurring from a perturbative analysis 
should be {\em anti-Hermitian}. 
A systematic formulation of this idea is to consider the
holomorphic matrix model instead.  
To substantiate this proposal, Appendix 2 shows
that the procedure of \cite{Klemm} admits a simple justification 
in the framework of holomorphic models.

\section{Loop equations, equations of motion and the large $N$ limit}

In this section, we study the loop equations and equations of motion of 
the holomorphic model, as well as its planar limit. As we shall see below, 
much of the standard fare of Hermitian models can be extended quite directly 
to the holomorphic case (though there are a few modifications). The 
most important novel
feature of holomorphic matrix models is that they can explore an enlarged 
set of planar limits, thus probing the entire 
moduli space of a certain family of algebraic curves. 
This fact,  which is essential for the conjecture of 
\cite{DV1}, will be established explicitly by proving a reconstruction 
theorem which associates a planar solution of the model with an arbitrary
algebraic curve belonging to this family.

Let $ds$ be the length element on $\gamma$ and $s$ the length coordinate 
along this curve, centered at some point on $\gamma$. We let $\lambda(s)$
denote the parameterization of $\gamma$ with respect to this coordinate
and write $\lambda_i=\lambda(s_i)$ accordingly. 
Mimicking usual constructions, we define the spectral density by:
\be
\rho(s)=\frac{1}{N}\sum_{j=1}^N{\delta(s-s_j)}~~,
\ee
where $\delta(s-s_j)$ is the delta-function in the coordinate $s$ 
(equivalently, the $\delta$-function along $\gamma$ with respect 
to the measure induced by the length element). 
Note the normalization condition:
\be
\label{rho_norm}
\int_{-\infty}^\infty{ds \rho(s)}=1~~.
\ee
Also consider the resolvent of $M$:
\be
R(z):=\frac{1}{z-M}~~,
\ee
and its normalized trace:
\be
\omega(z)=\frac{1}{N}
tr R(z)=\frac{1}{N}tr\frac{1}{z-M}=\frac{1}{N}
\sum_{i=1}^N{\frac{1}{z-\lambda_i}}=\int{ds \frac{\rho(s)}{z-\lambda(s)}}~~.
\ee

\begin{figure}[hbtp]
\begin{center}
\scalebox{0.6}{\input{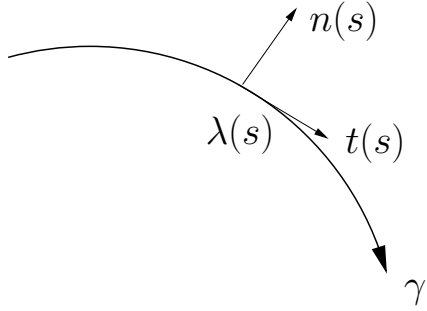}}
\end{center}
\caption{The normal vector field $n(s)=i t(s)$, 
and the tangent vector field $t(s)={\dot \lambda}(s)$ 
 of $\gamma$. Note that $|n(s)|=|t(s)|=1$ since $s$ is the length coordinate
 along $\gamma$.}
\label{normal}
\end{figure}

In the following, we shall need the Sokhotsky-Plemelj formulae:
\be
\label{Sokh}
\lim_{\epsilon \rightarrow 0^+}\int_\gamma
{d\lambda' \frac{1}{\lambda-\lambda'\pm \epsilon
    n(\lambda)}}={\cal P}\int_\gamma{d\lambda' \frac{1}{\lambda-\lambda'}}\mp
i\pi~~{\rm for}~~\lambda \in \gamma~~, 
\ee
which we also write symbolically as:
\be
\label{PV}
\lim_{\epsilon \rightarrow 0^+} \frac{1}{\lambda(s)-\lambda(s')\pm \epsilon
    n(s)}={\cal P} \frac{1}{\lambda(s)-\lambda(s')}\mp
\frac{i\pi}{{\dot \lambda}(s)}\delta(s-s')~~.
\ee
Here ${\cal P}$ stands for the principal value and $n(s)=
i {\dot \lambda}(s)$ is the normal
    vector field to $\gamma$ (figure \ref{normal}).

\subsection{Loop equations}

To extract the loop equations, we follow the method of \cite{Kostov}.
For this, start with the identity:
\be
\int_\gamma{d\lambda_1}\dots \int_\gamma{d\lambda_N}\sum_{i=1}^N{
\frac{\partial}{\partial \lambda_i}
\left[\prod_{k\neq l}{(\lambda_k-\lambda_l)} 
e^{-N\sum_{j=1}^N{W(\lambda_j)}}\frac{1}{z-\lambda_i}\right]}=0~~.
\ee
Performing the partial differentiation, we write this as:
\be
\langle 
\sum_{i=1}^N{\frac{1}{(z-\lambda_i)^2}}-
N\sum_{i=1}^N{\frac{W'(\lambda_i)}{z-\lambda_i}}+
2\sum_{i\neq j}{\frac{1}{(\lambda_i-\lambda_j)(z-\lambda_i)}}
\rangle=0~~.
\ee
Using the decomposition:
\be
\label{pf}
\frac{1}{(z-\alpha)(z-\beta)}=\frac{1}{\alpha-\beta}\left[
\frac{1}{z-\alpha}-\frac{1}{z-\beta}\right]
\ee
to simplify the last term and 
combining the result with the first gives:
\be
\label{loop_intmd}
\langle \omega(z)^2-\frac{1}{N}\sum_{i=1}^N{\frac{W'(\lambda_i)}{z-\lambda_i}}
\rangle=0~~.
\ee

Defining the polynomial:
\be
f(z):=\frac{1}{N}\sum_{i=1}^N{\frac{W'(z)-W'(\lambda_i)}{z-\lambda_i}}
=\int{ds \rho(s)\frac{W'(z)-W'(\lambda(s))}{z-\lambda(s)}}~~,
\ee
relation (\ref{loop_intmd}) gives the following form of the loop equations:
\be
\label{le}
\langle \omega(z)^2\rangle -W'(z)\langle \omega(z)\rangle+\langle f(z)
\rangle=0~~.
\ee

\subsection{Equations of motion}

Writing $\lambda_i=\lambda(s_i)$, the partition function (\ref{ev}) becomes:
\bea
{\cal Z}_N(\gamma,t)&=&\int{ds_1} \dots \int{ds_N}
\prod_{i=1}^N{\dot{\lambda}(s_i)}
\prod_{i\neq j}{(\lambda(s_i)-\lambda(s_j))} 
e^{-N\sum_{j=1}^N{W(\lambda (s_i))}}\nn\\
&=&\int{ds_1} \dots \int{ds_N}
e^{-N S_{eff}(s_1\dots s_N)}~~,
\eea
where $\dot{\lambda}(s):=\frac{d\lambda(s)}{ds}$ and:
\be
S_{eff}(s_1\dots s_N)=\sum_{j=1}^N{W(\lambda (s_i))}-\frac{1}{N}\sum_{i\neq
  j}{\ln(\lambda_i-\lambda_j)}-\frac{1}{N}\sum_{i=1}^N{\ln\dot{\lambda}_i}~~.
\ee
Extremizing with respect to $s_i$ gives the equations of motion:
\be
\label{eom}
\frac{2}{N}{\sum}'_{j}{\frac{1}{\lambda(s_i)-\lambda(s_j)}}=
W'(\lambda (s_i))-
\frac{1}{N}\frac{\ddot{\lambda}(s_i)}{\dot{\lambda}(s_i)^2}~~.
\ee
The prime means that the sum is taken only over $j\neq i$.
The last term is a curvature-induced contribution which is subleading in
$1/N$. It is also easy to check that the equations of motion imply:
\be
\omega(z)^2-W'(z)\omega(z)+f(z)+\frac{1}{N}\frac{d}{dz} \omega(z)+
\frac{1}{N^2}\sum_{i=1}^N{\frac{{\ddot \lambda}(s_i)}{{\dot \lambda}(s_i)^2}}
=0~~.
\ee
This `operator' equation holds only for solutions of (\ref{eom}), unlike the 
Ward identity (\ref{le}).

\subsection{The large $N$ limit}

\begin{figure}[hbtp]
\begin{center}
\scalebox{0.6}{\input{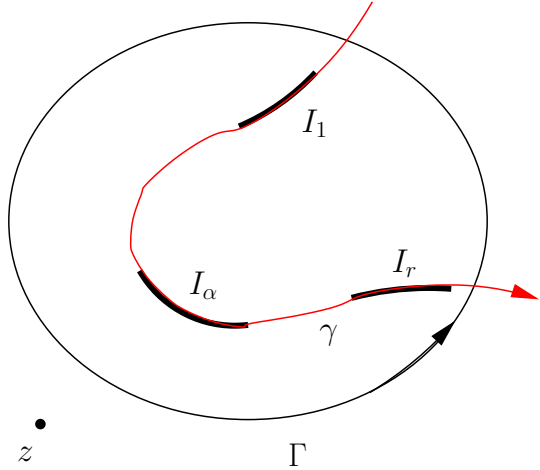}}
\end{center}
\caption{Cuts of $\omega_0$. We also show a closed 
contour $\Gamma$ surrounding all cuts and a point $z$ in its exterior. }
\label{cuts}
\end{figure}

For any quantity $\phi$, consider the large $N$ expansion:
\be
\label{phi_exp}
\langle \phi\rangle=\sum_{j=0}^\infty{\frac{\phi_j}{N^j}}~~,
\ee
with coefficients $\phi_0, \phi_1$ etc. In particular, we 
have $\langle \rho(s)\rangle =\rho_0(s)+O(1/N)$ and 
$\langle \omega(z)\rangle =\omega_0(z)+O(1/N)$. 
In the large $N$ limit, the eigenvalues $\lambda_j$ are replaced by the 
planar spectral density $\rho_0(\lambda)$ supported on the curve $\gamma$. 
Note that this quantity is {\em complex}, since its definition involves
averaging with respect to the complex integrand of (\ref{ev}). 
As usual, we have
\be
\label{omega0}
\omega_0(z)=
\int{ds\frac{\rho_0(s)}{z-\lambda(s)}}~~,
\ee
so that $\omega_0$ becomes an analytic function whose cuts $I_\alpha$ 
are forced to lie on the curve $\gamma$ (figure \ref{cuts}). 
We shall assume for simplicity that all cuts are of finite length.

In the planar limit, the average of a product can be replaced by the
product of averages and the loop equations (\ref{le}) reduce to the algebraic
constraint:
\be
\label{curve0}
\omega_0(z)^2 -W'(z)\omega_0(z)+f_0(z)=0~~,
\ee
where 
\be
\label{f0}
f_0(z)=
\int{ds\rho_0(s)\frac{W'(z)-W'(\lambda(s))}{
z-\lambda(s)}}
\ee
is a polynomial of degree $n-2$ with complex coefficients. 
Since $\rho(s)$ is normalized by (\ref{rho_norm}), 
equation (\ref{f0}) shows that the leading coefficient of 
$f_0(z)$ equals $nt_{n}$.

In the planar limit, the equations of motion (\ref{eom}) become:
\be
\label{eom0}
2{\cal P}\int{ds'}{\frac{\rho_0(s')}{\lambda(s)-\lambda(s')}}=W'(\lambda(s))~~.
\ee
The curvature term in (\ref{eom}) drops out, since it is 
subleading in $1/N$. 
The algebraic constraint (\ref{curve0}) can also be obtained from (\ref{eom0})
by standard manipulations. 
\noindent Writing:
\be
\label{shift}
\omega_0(z)=u_0(z)+\frac{1}{2}W'(z)~~, 
\ee
relation (\ref{curve0}) 
shows that $u_0(z)$ is one of the branches of the planar affine curve: 
\be
\label{curve}
u^2-\frac{1}{4}W'(z)^2+f_0(z)=0~~. 
\ee

Since the branch cuts of $\omega_0$ (and thus of $u_0$) 
must lie along 
$\gamma$, it is clear that the polynomial $f_0$ in equation (\ref{curve}) 
is constrained by the choice of this curve. 
However, changing $\gamma$ without changing its asymptotes allows one
to describe any position of the cuts in the complex plane, 
as long as all these cuts have finite length. 
This effectively eliminates the constraints that would be present in 
the Hermitian case (for which $\gamma$ would be forced to coincide with the
real axis).

\subsection{Reconstruction of a planar solution from the Riemann surface}

Given the algebraic curve (\ref{curve}), we now show how one can use it to 
recover an appropriate $\gamma$ supporting a 
spectral density $\rho_0$ satisfying (\ref{rho_norm}), (\ref{omega0}), 
(\ref{f0}) and the planar equations of motion (\ref{eom0}). 
This proves that the holomorphic matrix model is free to explore the 
whole relevant piece of the moduli space of (\ref{curve}), 
unlike the Hermitian matrix model. In the planar limit, the holomorphic 
model reduces to the singular integral equation (\ref{eom0}). 
Up to some minor details, 
the existence of a one to one relation between solutions of this 
equation and members of a family of Riemann surfaces boils down to the
well-known relation between the Riemann problem and 
singular integral equations.

To see this explicitly, assume that one is given a complex degree 
$n-2$ polynomial $f_0(z)$, subject only to the constraint that its leading
coefficient equals $n t_n$. Consider the associated  curve (\ref{curve}).
Denoting its branch points by $a_\alpha, b_\alpha$ (with 
$\alpha=1\dots n-1$), choose branch-cuts $I_\alpha$  
connecting $a_\alpha$ and $b_\alpha$ (note that we allow $I_\alpha$ 
to be curved). This defines 
two determinations, which we call $u_0$ and $u_1=-u_0$. More precisely, $u_0$ 
is the determination which behaves as $-\frac{1}{2}W'(z)$ for large $z$.
Let us define $\omega_0$ by relation (\ref{shift}) 
and choose a curve $\gamma$ such that $I_\alpha\subset \gamma$ for all
$\alpha$. We let $s$ be its length coordinate and 
$\lambda=\lambda(s)$ the associated parameterization.
Choosing the normal $n(s)=i{\dot \lambda}(s)$ (figure \ref{normal}) we define:
\be
\label{rho0_def}
\rho_0(s):={\dot \lambda}(s) \lim_{\epsilon\rightarrow 0^+}\frac{1}{2\pi i}
\left[\omega_0(\lambda(s)-\epsilon n(s))-
\omega_0(\lambda(s)+\epsilon n(s))\right]~~
\ee
for every $s$ in $\gamma$. Then $\rho$ vanishes outside of $I_\alpha$
and (\ref{rho0_def}) and the Sokhotsky formulae (\ref{Sokh}) imply:
\be
\int{ds \frac{\rho_0(s)}{z-\lambda(s)}}=\int_\Gamma{\frac{dx}{2\pi i}
\frac{\omega_0(x)}{z-x}}=\omega_0(z)~~,
\ee
where $\Gamma$ is a contour surrounding all cuts but not the point $z$ 
(see figure \ref{cuts})
and the last equality follows by deforming this contour toward infinity 
to pick the contribution from the residue at $x=z$. This shows that 
relation (\ref{omega0}) holds. Using (\ref{curve}) and the fact that 
the leading coefficient of $f_0$ equals $n t_n$ shows that 
$\omega(z)=1/z+O(1/z^2)$ for large $|z|$, and combining this with 
(\ref{omega0}) shows that $\rho_0$ satisfies the normalization condition
(\ref{rho_norm}). 

Since the cuts $I_\alpha$ connect the branches 
$u_0$ and $-u_0$, we have:
\be
\lim_{\epsilon\rightarrow 0^+}{u_0(\lambda +\epsilon n )}=
\lim_{\epsilon\rightarrow 0^+}{[-u_0(\lambda - \epsilon n )
]}~~,~~\lambda \in I_\alpha
\ee
so that $\lim_{\epsilon \rightarrow 0^+}\left[
\omega_0(\lambda+\epsilon n)+\omega_0(\lambda-\epsilon n)\right]=W'(\lambda)$
for $\lambda \in I_\alpha$.
Combining this with equation (\ref{omega0}) and using the
Sokhotsky formulae (\ref{Sokh}) shows that $\rho_0$ 
satisfies the planar equations of motion (\ref{eom0}) along the cuts.

To prove (\ref{f0}), we use relation (\ref{omega0}) to compute:
\be
\label{int}
\omega_0(z)^2=\int{
ds \int
{ds'\frac{\rho_0(s)\rho_0(s')}{(z-\lambda(s))(z-\lambda(s'))}}}=
2\int{
ds\int{ds' 
\frac{\rho_0(s)\rho_0(s')}{(\lambda(s)-\lambda(s'))(z-\lambda(s))}}}~~,
\ee
where we used the identity (\ref{pf})
and symmetry of the resulting integrand with respect to the substitution 
$s\leftrightarrow s'$. Using 
(\ref{eom0}) in the right hand side of (\ref{int}) gives:
\be
\omega_0(z)^2=\int{ds\rho_0(s)\frac{W'(\lambda(s))}{
z-\lambda(s)}}=-\int{ds\rho_0(s)\frac{W'(z)-W'(\lambda(s))}{
z-\lambda(s)}}+W'(z)\omega_0(z)~~.
\ee
Comparing with (\ref{curve}) and (\ref{shift}) 
shows that equation (\ref{f0}) holds. 

This construction produces a planar solution of the holomorphic
model which recovers {\em any} curve of the form (\ref{curve}), for an 
arbitrary choice of degree $n-2$ polynomial $f_0(z)$ with leading coefficient 
$nt_n$. Unlike the Hermitian model, 
the holomorphic model explores the entire family of curves 
(\ref{curve}) in its planar limit. This is a basic pre-requisite for 
the conjecture of \cite{DV1}.

It is also clear that the precise choice of cuts 
$I_\alpha$ is irrelevant, as long as they connect the given pairs of 
branch points $a_\alpha,b_\alpha$. Thus one can use 
any\footnote{Nondegenerate and without self-intersections.} 
curve $\gamma$ in this construction, provided that it 
passes through all the branch points of (\ref{curve}).

\section{The microcanonical ensemble}

The framework of \cite{DV1} requires that the model obeys 
certain filling fraction constraints. 
In this section, I explain how one can impose such constraints on the 
{\em finite $N$} model\footnote{A clear formulation of such constraints 
beyond the strict large $N$ limit is important in studies of
$SO(N)$ and $Sp(N)$ matrix models, whose large $N$ expansion contains 
terms of order $1/N$ which must be interpreted as contributions to the 
dual field theory (rather than gravitational contributions, which start at 
order $1/N^2$). The formulation given below allows one to give 
clear proofs of relations between the strictly planar and $O(1/N)$ 
contributions to the (microcanonical) partition function, thereby 
strengthening 
arguments like those given in \cite{Corrado}. Similar problems are encountered 
in more general orientifold models \cite{us}.}. The relevant 
conditions are easiest to formulate by employing a
microcanonical ensemble. As we shall see below, 
the original path integral defines a 
(grand-)canonical ensemble at zero chemical potentials. This 
allows one to recover the microcanonical generating function by introducing
chemical potentials (which are canonically conjugate to the filling fractions)
and then performing a Legendre transform to replace the former by the latter.

\subsection{The (grand-)canonical partition function associated with a 
collection of domains}

\begin{figure}[hbtp]
\begin{center}
\scalebox{0.5}{\input{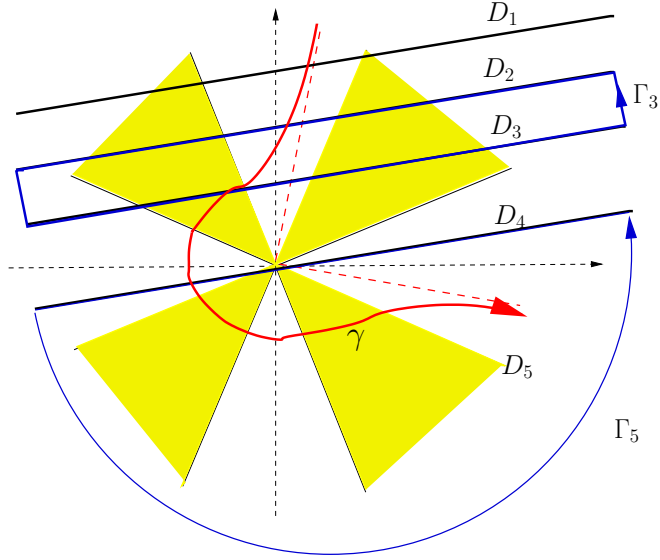}}
\end{center}
\caption{A choice of strip domains in the complex plane. 
We also show two of the 
bounding contours (namely $\Gamma_3$ and $\Gamma_5$). In this example, we take
the domains to be infinite strips; this assures us that $\gamma$ and 
its deformations 
will cut each such domain along a {\em non-void} interval. }
\label{domains}
\end{figure}

We start by fixing a collection of domains $D_\alpha$ ($\alpha=1\dots r$) 
in the complex plane, chosen such that their interiors are mutually disjoint 
and such that:
\be
\label{part}
\cup_{\alpha=1}^r{{\overline D}_\alpha}=\C~~.
\ee
We shall assume that $\gamma$ 
intersects each closure ${\bar D}_\alpha$ along a single curve segment 
$\Delta_\alpha$, where $\Delta_\alpha$ lie 
in ascending order on $\gamma$ with respect to its 
orientation (figure \ref{domains}). Condition (\ref{part}) implies:
\be
\label{partition}
\cup_{\alpha=1}^r{\Delta_\alpha}=\gamma~~.
\ee 
For simplicity, we shall take $D_\alpha$ to be 
infinite strips arranged as in figure \ref{domains} . 

Letting $\chi_\alpha$ denote the characteristic function of 
$D_\alpha$, consider the matrices
$\chi_\alpha(M)$ defined by holomorphic functional calculus:
\be
\chi_\alpha(M)=\oint_{\Gamma_\alpha}{\frac{dz}{2\pi i}\frac{1}{z-M}}~~,
\ee
with $\Gamma_\alpha$ a (counterclockwise) contour bounding 
$D_\alpha$ in the complex plane (figure \ref{domains}).  
The matrix $\chi_\alpha(M)$ equals the projector on the space spanned by those 
eigenvectors of $M$ whose eigenvalues lie in $D_\alpha$. 

Let $f_\alpha:=\frac{1}{N}tr\chi_\alpha(M)=
\frac{1}{N}\sum_{j=1}^N{\chi_\alpha(\lambda_j)}$ be the 
filling fractions for the domains $D_\alpha$. One has:
\be
f_\alpha
=\int{ds\rho(s)\chi_\alpha(\lambda(s))}=
\oint_{\Gamma_\alpha}{\frac{dz}{2\pi i}\omega(z)}~~,
\ee
since $\omega(z)$ has simple poles with residue $\frac{1}{N}$ 
at each eigenvalue $\lambda_j$. 
Relation (\ref{partition}) implies $\sum_{\alpha=1}^r{\chi_\alpha(\lambda)}=1$,
so the filling fractions are subject to the constraint:
\be
\label{norm}
\sum_{\alpha=1}^r{f_\alpha}=1~~.
\ee

Picking {\em complex} chemical potentials $\mu_\alpha$, we consider the 
(grand-)canonical ensemble associated with our collection of domains:
\bea
{\cal Z}_N(\gamma;t,\mu)&=&\frac{1}{{\cal N}}
\int_{\Gamma(\gamma)}{ dM e^{tr\left[ -N W(M)-
N\sum_{\alpha=1}^r{\mu_\alpha\chi_\alpha(M)}\right]}}\nn\\
&=&\int_\gamma{d\lambda_1 \dots \int_\gamma d\lambda_N}
\prod_{i\neq j}{(\lambda_i-\lambda_j)} e^{-N\sum_{j=1}^N{W(\lambda_j)}-
N^2\sum_{\alpha=1}^r{\mu_\alpha f_\alpha}}~~,
\eea
which is an analytic function of $\mu_\alpha$. 
The original partition function results by setting 
$\mu_\alpha=0$, and thus it corresponds to a {\em (grand-)canonical}
ensemble at zero chemical potentials. Notice that the 
(grand-)canonical partition function can be written:
\be
{\cal Z}_N(\gamma;t,\mu)=\sum_{N_1+\dots +N_r=N}{\frac{N!}{N_1!\dots N_r!}}
e^{-N\sum_{\alpha=1}^N{\mu_\alpha N_\alpha}}
{\cal Z}_{N_1\dots N_r}(\gamma;t)~~,
\ee
where:
\be
{\cal Z}_{N_1\dots N_r}(\gamma;t)=
\int_\gamma{d\lambda_1 \dots \int_\gamma d\lambda_{N_1}}
\dots
\int_\gamma{d\lambda_{N_1+\dots +N_{r-1}+1} 
\dots \int_\gamma d\lambda_{N}}
\prod_{i\neq j}{(\lambda_i-\lambda_j)} e^{-N\sum_{j=1}^N{W(\lambda_j)}}~~.
\ee

{\bf Observation} One can consider a more general version of microcanonical 
ensemble based on a finite partition of unity, i.e. a finite collection
of smooth complex-valued 
functions $\phi_\alpha(z,{\overline z})$ satisfying the 
constraint:
\be
\sum_{\alpha}{\phi_\alpha}=1~~.
\ee
In certain ways, this is preferable to the approach taken above, since 
it may be technically desirable to avoid having to deal with characteristic 
functions. The entire discussion of this section extends easily to this 
more general set-up.

Introducing the (grand-)canonical generating function:
\be
{\cal F}_N(\gamma;t,\mu):=-\frac{1}{N^2}\ln {\cal Z}_N(\gamma;t,\mu)~~,
\ee
we have the standard relation:
\be
\label{cfractions}
\frac{\partial}{\partial \mu_\alpha}{\cal F}=\langle f_\alpha\rangle~~.
\ee
Here and below, the brackets $\langle \dots \rangle$ denote the 
expectation value taken in the {\em (grand-)canonical} ensemble.

\subsection{The microcanonical generating function}

Following standard statistical mechanics procedure, we define: 
\be
\label{S_def}
S_\alpha:=\frac{\partial}{\partial\mu_\alpha}{\cal F}
\ee
and perform a Legendre transform to extract the microcanonical generating 
function:
\be
\label{F_def}
F(\gamma, t,S):=S_\alpha\mu_\alpha(\gamma, t,S) -
{\cal F}(\gamma, t,\mu_\alpha(t, S))~~,
\ee
which is an analytic function of the complex variables $t_\alpha$ and
$S_\alpha$. 
In this relation, $\mu_\alpha$ are expressed in terms of $t$ and $S$ by
solving equations (\ref{S_def}). The constraint (\ref{norm}) 
and equation (\ref{cfractions}) show that 
$S_\alpha$ satisfy:
\be
\label{Snorm}
\sum_{\alpha=1}^N{S_\alpha}=1~~,
\ee
so we can take $S_1\dots S_{r-1}$ to be the independent variables. Then 
equations (\ref{S_def}) express $\mu_\alpha$ as functions of $t_j$ and
these coordinates, and equation (\ref{F_def}) implies:
\be
\label{quantum_periods}
\mu_\alpha-\mu_r=\frac{\partial F}{\partial S_\alpha}~{\rm~for~}\alpha=
1\dots r-1~~.
\ee
This  gives the chemical potentials as functions of $t$ and $S_\alpha$. 
Note that $\mu_\alpha$ are only determined up to a common constant shift; 
this is due to the constraint (\ref{Snorm}) on $S_\alpha$. 
Working with $F(\gamma, t,S)$ amounts to fixing the expectation 
values of the filling fractions by imposing the {\em quantum} constraint
(\ref{S_def}):
\be
\label{fconstraint}
\langle f_\alpha \rangle=
\oint_{\gamma_\alpha}{\frac{dz}{2\pi i}\langle \omega(z)\rangle}=S_\alpha~~,
\ee
with $S_\alpha$ treated as fixed parameters. We stress that this condition
is only imposed on the expectation values 
of the filling fractions.

\subsection{Chemical potentials at large $N$}

We will now show that the large $N$ 
chemical potentials can be expressed as B-type
periods of the algebraic curve (\ref{curve}), thereby proving 
that the special geometry relations of \cite{DV1} hold at the holomorphic 
matrix level. The argument below is an adaptation of that given in
\cite{Whitham}\footnote{In \cite{DV1}, Dijkgraaf and Vafa 
gave a beautiful intuitive
  justification for the special geometry relation $\Pi_\alpha=\frac{\partial
    F_0}{\partial S_\alpha}+const$, where $F_0$ is the planar limit of the 
(microcanonical) generating 
function while
$S_\alpha$ and $\Pi_\alpha$ are
identified with periods of the curve (\ref{curve}). A  
derivation of this 
relation (in the context of the Hermitian model) was later given in
\cite{Whitham}, upon using older results of \cite{Jurk}.}, 
combined with the definition of the microcanonical ensemble
given above. In particular, we show that the special geometry relations are 
simply the large $N$ limit of the standard equations (\ref{quantum_periods}).
Hence the chemical potentials $\mu_\alpha$ are 
the `quantum' (i.e. finite $N$) analogues of the B-type periods of \cite{DV1}, 
while the averaged filling fractions are the `quantum' analogues of the
$A$-type periods. This captures the beautiful intuition of \cite{DV1}.

For this, we start from the expression of the (grand-)canonical 
generating function in the planar limit:
\be
\label{F_can}
{\cal F}_0(\gamma, t,\mu)=\int{ds W(\lambda(s))\rho_0(s)}-
{\cal P}\int{ds}\int{ds'
{\tilde K}(\lambda(s),\lambda(s'))\rho_0(s)
\rho_0(s')}+
\sum_{\alpha=1}^r{\mu_\alpha\int_{I_\alpha}{ds\rho_0(s)}}~~
\ee
where:
\be
{\tilde K}(\lambda,\lambda'):=\ln(\lambda-\lambda')~~.
\ee
In what follows, we shall assume that each cut $I_\alpha$ 
lies inside a 
corresponding domain $D_\alpha$; in particular, we assume that the number of
cuts coincides with the number of domains.  
With this assumption, we have
$S_\alpha=\langle f_\alpha\rangle=
\int_{I_\alpha}{ds\rho_0(s)}$ and
the Legendre transform of (\ref{F_can}) 
gives the planar limit of the microcanonical generating function:
\be
\label{F_0}
F_0(\gamma, t,S)={\cal P}\int
{ds}\int{ds'{\tilde K}(\lambda(s),\lambda(s'))
\rho_0(s)\rho_0(s')}-\int
{ds W(\lambda(s))\rho_0(s)}~~,
\ee
with the constraints:
\be
\label{csrt}
\int_{I_\alpha}{ds\rho_0(s)}=S_\alpha~~
{\rm for}~~\alpha=1\dots r~~.
\ee
Remember that we 
allow $I_\alpha$ to be curved intervals connecting the branch points
$a_\alpha$ and $b_\alpha$ of the algebraic curve (\ref{curve}). In the 
generic case, none of the cuts is reduced to a double point and one can take 
$r=n-1$.

\subsubsection{The primitive of $u_0$ along $\gamma$}

To extract the large $N$ chemical potentials, 
consider the `restriction' of $u_0$ along $\gamma$, which
we define by:
\be
u_0^p(s):=
\frac{1}{2}\lim_{\epsilon \rightarrow 0^+}
\left[u_0(\lambda(s)+\epsilon n(s))+u_0(\lambda(s)-\epsilon n(s))
\right]~~.
\ee
Here $n(s)=i{\dot \lambda}(s)$.

If $\lambda(s)$ is a point of $\gamma$ lying outside the union of 
$I_\alpha$, then $u_0^p(s)$ equals $u_0(\lambda(s))$, the quantity
obtained by substituting $\lambda(s)$ for $z$ in the expression:
\be
\label{yu}
u_0(z)=\omega_0(z)-\frac{1}{2}W'(z)
=\int{ds' \frac{\rho_0(s')}{z-\lambda(s')}}-\frac{1}{2}
W'(z)~~.
\ee

Using (\ref{yu}) in the definition of $u_0^p$ gives:
\be
\label{u_0p}
u_0^p(s):={\cal P}\int{ds' \rho_0(s')K(\lambda(s),\lambda(s'))}-
\frac{1}{2}W'(\lambda(s))~~,
\ee
where:
\be
K(\lambda,\lambda')=\frac{1}{\lambda-\lambda'}
\ee
is the integral kernel appearing in the planar loop equations 
and where we used the Sokhotsky identities (\ref{Sokh}). 

Consider now the function $\phi:\gamma\rightarrow \C$ defined though:
\be
\label{phi_def}
\phi(s):=2{\cal P}\int{ds'{\tilde K}(\lambda(s),\lambda(s'))
\rho_0(s')}-W(\lambda(s))~~.
\ee
Noticing that $K(\lambda,\lambda')=\frac{\partial }{\partial 
\lambda}{\tilde K}(\lambda,\lambda')$, we have:
\be
\label{dphi}
\frac{d}{ds}\phi(s)=2\dot{\lambda}(s)u_0^p(s)~~.
\ee

As clear from (\ref{u_0p}), the planar
equations of motion (\ref{eom0}) amount to the requirement that
$u_0^p$ vanishes along each of the curve segments $I_\alpha$:
\be
u_0^p(s)=0~~{\rm for}~~\lambda(s) \in I:=
\cup_{\alpha=1}^r{I_\alpha}~~.
\ee
This means that $\phi$ is constant along each of these intervals:
\be
\label{constancy}
\phi(s)=\xi_\alpha={\rm~constant~on~}I_\alpha~~,~~
\ee
The jump in the value of $u_0$ between 
consecutive cuts can be obtained by integrating (\ref{dphi}):
\bea
\label{xi_int0}
\xi_{\alpha+1}-\xi_\alpha &=&
2\int_{b_{\alpha}}^{a_{\alpha+1}}{d\lambda u_0(\lambda)}~~,
\eea
where we used $d\lambda=\dot{\lambda}(s)ds$.
This integral is of course taken along $\gamma$.

\subsubsection{The planar chemical potentials}

Differentiating (\ref{F_0}) with respect to $S_\alpha$ for some $\alpha<r$ and
using relation (\ref{phi_def}) gives:
\be
\label{mu_xi}
\mu^{(0)}_\alpha-\mu_r^{(0)}=\frac{\partial}{\partial S_\alpha}
F_0(\gamma, t,S)=
\int_{\cup_{\beta=1}^r{I_\beta}}
{ds \frac{\partial \rho_0(s)}{\partial S_\alpha}\phi(s)}=
\xi_\alpha-\xi_r~~,
\ee
where $\mu^{(0)}_\alpha$ are the planar limits of the chemical potentials. 
To arrive at the last equality, we used equation (\ref{constancy}) and 
noticed that 
$\int_{I_\beta}\frac{\partial \rho_0(\lambda)}{\partial S_\alpha}=
\frac{\partial }{\partial S_\alpha}\int_{I_\beta}{ds\rho_0(s)}$
equals $\delta_{\alpha\beta}$ if $\beta<r$  
and $-1$ if $\beta=r$ (by virtue of (\ref{csrt}) and (\ref{Snorm})). 
Relation (\ref{mu_xi}) shows that the planar chemical potentials coincide with 
the quantities $\xi_\alpha$, up to a common 
additive constant. Using relation (\ref{xi_int0}), we obtain:
\be
\label{xi_int1}
\mu^{(0)}_{\alpha+1}-\mu^{(0)}_\alpha=
2\int_{b_{\alpha}}^{a_{\alpha+1}}{d\lambda u_0(\lambda)}
\ee

\begin{figure}[hbtp]
\begin{center}
\scalebox{0.5}{\input{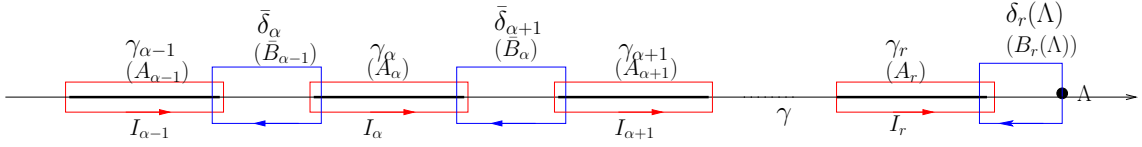}}
\end{center}
\caption{Choice of A and B-cycles on the large $N$ Riemann surface 
(for convenience, we represent the curve $\gamma$ as a straight line, though
this need not be the case).  The figure indicates the projections
$\gamma_\alpha$, ${\bar \delta}_\alpha$ and $\delta_r(\Lambda)$ 
of the cycles $A_\alpha$, ${\bar B}_\alpha$ and $B_r$ onto the $z$ plane.  
The cycles $A_\alpha$ can be identified with their 
projections $\gamma_\alpha$.
The cycles ${\bar B}_\alpha$ are defined such that, when
crossing the cut $I_\alpha$ going upwards along these cycles, one moves
from the branch $u_0$ to the branch $u_1=-u_0$. 
Thus the `lower halves' of these
cycles lie on the branch $u_0$, while their `upper halves' lie on the branch
$u_1$. A similar convention is used for $B_r(\Lambda)$.
With the orientation of the Riemann surface induced by its complex
structure, this implies the intersections $A_\alpha \cap {\bar B}_\alpha=+1$
(note that the cycles $A_\alpha$ and ${\bar B}_\alpha$ intersect in a single
point, which lies on the branch $u_0$).}
\label{Bcycles}
\end{figure}

As explained above, the quantity $u_0(z)$ has cuts precisely along 
the intervals $I_\alpha$. Since the other branch of 
(\ref{curve}) is given by $u_1(z)=-u_0(z)$, this 
allows us to write (\ref{xi_int1}) in the form:
\be
\label{xi_int}
\mu^{(0)}_{\alpha}-\mu^{(0)}_{\alpha+1}=
\int_{b_{\alpha}}^{a_{\alpha+1}}{dz [u_1(z)-u_0(z)]}
=\oint_{{\bar B}_\alpha}{dz u(z)}~~,
\ee
where ${\bar B}_\alpha$ are cycles on the large $N$ Riemann surface 
chosen as explained in figure \ref{Bcycles}.

\

Consider the cycles 
${\tilde B}_\alpha=\sum_{\beta=\alpha}^{r-1}{\bar B}_\beta$ 
for all $\alpha=1\dots r-1$. Then (\ref{xi_int}) implies:
\be
\mu^{(0)}_\alpha=\mu^{(0)}_r+
\oint_{{\tilde B}_\alpha}{dz u(z)}~~{\rm~for~}~~\alpha=1\dots r-1~~.
\ee
The quantity $\mu_r^{(0)}$ is undetermined and can be fixed arbitrarily. 
Following \cite{DV1}, we take $\mu_r^{(0)}=\oint_{B_r}{dz u(z)}$, 
where $\Lambda$ is a point close to infinity and 
$B_r(\Lambda)$ is the cycle described in figure \ref{Bcycles}. Defining 
$B_\alpha(\Lambda)={\tilde B}_\alpha+B_r(\Lambda)$ 
for all $\alpha=1\dots r-1$,  we obtain:
\be
\label{mu_Pi}
\mu^{(0)}_\alpha=\Pi_\alpha {\rm~~for~~}\alpha=1\dots r~~,
\ee
with:
\be
\label{Pi}
\Pi_\alpha:=\int_{B_\alpha}{dz u(z)}~~.
\ee
Relation (\ref{mu_Pi}) shows that the chemical potentials 
$\mu_\alpha$ are the finite $N$ analogues of the periods $\Pi_\alpha$. 

Note also that the filling fractions can be expressed as periods of 
the meromorphic 
differential $udz$ over the cycles $A_\alpha$ of figure \ref{Bcycles}:
\be
S_\alpha=\oint_{\gamma_\alpha}{\frac{dz}{2\pi i}
\omega_0(z)}=\oint_{\gamma_\alpha}{\frac{dz}{2\pi i}u_0(z)}=
\oint_{A_\alpha}{\frac{dz}{2\pi i}u(z)}~~.
\ee
In the second equality, we used relation (\ref{shift}) and the fact that 
$W(z)$ is a polynomial. Equation (\ref{quantum_periods}) now gives the special
geometry relation of \cite{DV1}:
\be
\Pi_\alpha-\Pi_r=\frac{\partial }{\partial S_\alpha}F_0~~.
\ee
Note that the quantity in the right hand side is the 
planar limit of the {\em microcanonical} generating function.

It is clear from figure \ref{Bcycles} that 
$A_\alpha\cap {\bar B}_\alpha=
-A_\alpha\cap {\bar B}_{\alpha-1}=+1$ and $A_\alpha\cap {\bar B}_\beta=0$ 
if $\beta\neq \alpha, \alpha-1$. This gives 
$A_\alpha\cap B_\beta=-B_\alpha\cap A_\beta=\delta_{\alpha\beta}$.
Since we also have $A_\alpha\cap A_\beta=B_\alpha\cap B_\beta=0$, it follows
that $A_\alpha,B_\beta$ have canonical intersection form. 
Hence we have a canonical system of cycles $A_\alpha,B_\alpha$ with 
$\alpha= 1\dots r$ :
\be
A_\alpha\cap B_\beta=-B_\alpha\cap A_\beta=\delta_{\alpha,\beta}~~,~~
A_\alpha\cap A_\beta=B_\alpha\cap B_\beta=0~~{\rm for~all}~~
\alpha=1\dots r~~.
\ee
This shows that the properties essential for  the conjecture of 
\cite{DV1} hold at the 
level of the holomorphic matrix model. 

\section{Holomorphic ADE models}

In this section, we consider the case of holomorphic $ADE$ models, 
focusing on the simple example of the holomorphic $A_2$ model. 
As mentioned in the introduction, the Hermitian approach to such models 
(and its attending regularization, discussed in \cite{Kostov_ADE}) leads 
to various technical problems which would violate the conjecture of 
\cite{DV1}. The purpose of this section is to show explicitly how 
such issues are avoided in the holomorphic set-up, and to provide 
a reconstruction theorem similar to the one found for the one-matrix 
case. In particular, we shall extract explicitly 
the associated Riemann surface (which is expected from the 
work of of \cite{DV3})
and show that one must use a certain `renormalization' procedure in order 
eliminate unwanted constraints on its moduli. In fact, we shall find 
that the curve expected from the work of \cite{DV3} and 
\cite{Cachazo} can be obtained by using a certain regulator 
(which is natural in the holomorphic set-up), though only in the limit where 
this regulator is removed. As we shall see below, this limit can be described 
as a statistical ensemble of `reduced' holomorphic models.

\subsection{Construction of the models}

Consider an $ADE$ quiver diagram with nodes indexed by $\alpha=1\dots \kappa $ 
where 
$\kappa $ is the rank of the associated simply-laced group. Consider 
$N_\alpha\times N_\alpha$ complex matrices $\Phi^{(\alpha)}$ 
for each node, and 
$N_\alpha \times N_\beta$ complex 
matrices $Q^{(\alpha\beta)}$ for each pair of
nodes $\alpha,\beta$ which are connected by an edge (in particular, 
we have {\em two} matrices $Q^{(\alpha\beta)}$ and 
$Q^{(\beta\alpha)}$ 
for each edge of the quiver). We let $s_{\alpha\beta}=s_{\beta\alpha}$ be the 
incidence matrix of the quiver and 
$c_{\alpha\beta}=2\delta_{\alpha\beta}-s_{\alpha\beta}$ be the 
associated Cartan matrix. 

By analogy with \cite{Kostov_ADE}, we define
the {\em holomorphic ADE matrix model} associated with this quiver by:
\be
\label{ZADE}
{\cal Z}=\int_{\Gamma}{\prod_{\alpha=1}^\kappa
{d\Phi^{(\alpha)} \prod_{\alpha<\beta}{\left
[dQ^{(\alpha\beta)}dQ^{(\beta\alpha)}\right]
e^{-N\sum_{\alpha=1}^\rho {W_\alpha(\Phi^{(\alpha)})}+W_{int}(\Phi,Q)}
}}}~~,
\ee
where $W_\alpha$ are polynomials of degrees $n_\alpha$ and:
\be
W_{int}(\Phi,Q):=\sum_{\alpha <\beta}{s_{\alpha\beta}
\left[tr(Q^{(\alpha\beta)}\Phi^{(\beta)}Q^{(\beta\alpha)})-
tr(Q^{(\beta\alpha)}\Phi^{(\alpha)}Q^{(\alpha\beta)})\right]}~~.
\ee
The gauge group is:
\be
G:=\prod_{\alpha=1}^\kappa {GL(N_\alpha,\C)}~~,
\ee
with the the obvious action:
\be
\label{Gaction}
(\Phi^{(\alpha)}, Q^{(\alpha\beta)})\rightarrow 
(S_\alpha \Phi^{(\alpha)}S_{\alpha}^{-1}, 
S_{\alpha}Q^{(\alpha\beta)}S_\beta^{-1})
\ee
for $S_\alpha\in GL(N_\alpha,\C)$.

To completely specify the model, one must choose an appropriate integration 
manifold $\Gamma$. Before explaining our choice, let us comment on the 
Hermitian approach \cite{Kostov_ADE}. In that case, one 
takes $\Gamma$ to consists of matrices $\Phi^{(\alpha)},Q^{(\alpha\beta)}$ 
such that $\Phi^{(\alpha)}$ are Hermitian and 
$Q^{(\beta\alpha)}=(Q^{(\alpha\beta)})^\dagger$ 
for neighboring nodes $\alpha$ and $\beta$.
Such a prescription makes sense only
if all $W_\alpha$ have even degree (otherwise, the integral diverges 
because the absolute value of the integrand explodes when the norm of some
$\Phi^{(\alpha)}$ is large). 
However, the Hermitian prescription immediately leads 
to other problems, arising 
from the integrals over $Q$. Indeed, it is easy to see 
that these will bring infinite contributions 
when some
eigenvalue $\lambda^{(\alpha)}_i$ of $\Phi^{(\alpha)}$ coincides with 
some eigenvalue 
$\lambda^{(\beta)}_j$ of $\Phi^{(\beta)}$ for neighboring $\alpha$ and $\beta$.
In the Hermitian framework,
the solution to this problem is to require
that the eigenvalues of neighboring $\Phi^{(\alpha)}$ 
can never coincide --- for example, by taking $\Phi^{(\alpha)}$ 
to have alternately negative and positive eigenvalues
\cite{Kostov_ADE}. It turns out that this prescription would violate the
conjecture of \cite{DV1}. To understand why, consider for simplicity the
Hermitian $A_2$ model (whose holomorphic version is studied below).
The large $N$ limit of this Hermitian model 
can be extracted by an argument which is formally identical to that presented 
in Appendix 3\footnote{Except that the third equation in (\ref{A2eom0}) 
used in the appendix never plays any role for the regularization of 
\cite{Kostov_ADE}.}, 
and is governed by an algebraic curve which is a triple 
cover of the complex plane. Denoting its three branches by $u_1$, $u_2$ and 
$u_3$ (in a convenient enumeration), one finds that cuts connecting $u_1$
and $u_2$ would correspond precisely to loci where {\em equal} eigenvalues
of $\Phi^{(1)}$ and $\Phi^{(2)}$ accumulate --- a situation which is forbidden 
by the regularization prescription used to define the model ! Therefore,
the regularization prescription of the Hermitian model 
requires that all such cuts are reduced to
double points, which means that the large $N$ curve must 
always be singular. Moreover, this violates the
requirements of the dual field theory, since it would require that some 
filling fractions vanish identically, thereby violating the study of gaugino
condensates performed in \cite{Cachazo}. Of course, the nonzero cuts of the 
resulting curve would also be constrained to alternately lie on the positive 
and negative halves of the real axis. 
It should be clear from this discussion that 
Hermitian $ADE$ models are quite unnatural  for the conjecture of \cite{DV1}.
Below, we shall show how a certain regularization prescription of the 
holomorphic $A_2$ model allows one to obtain an ensemble whose 
large $N$ limit satisfies the basic requirements of the conjecture. 
Unlike the Hermitian regularization of \cite{Kostov_ADE}, the `holomorphic'
regulator used below can be removed in a manner which allows us to recover 
a smooth Riemann surface. 

Returning to the holomorphic model, 
we shall choose the integration manifold $\Gamma$ as follows:  

(1) Fix contours 
$\gamma^{(\alpha)}$ in the complex plane such that:
\be
\label{reg}
\gamma^{(\alpha)}\cap \gamma^{(\beta)}=\emptyset~~{\rm~~for~neighboring}~~
\alpha~~{\rm and}~~\beta~~.
\ee
and such that each $\gamma^{(\alpha)}$ connects two convergence sectors of
$W_\alpha$ (as defined in Section 2). 

(2) Let $\Delta(\gamma)$ be the set of all matrices 
$(D^{(\alpha)}, Q^{(\alpha\beta)})$ (with $\alpha, \beta=1\dots \kappa $)
which satisfy: 

(a)$D^{(\alpha)}={\rm diag}
(\lambda^{(\alpha)}_1\dots \lambda^{(\alpha)}_{N_\alpha})$, with {\em distinct}
$\lambda^{(\alpha)}_1\dots \lambda^{(\alpha)}_{N_\alpha}$.

(b)$\lambda^{(\alpha)}_j$ lies along 
$\gamma^{(\alpha)}$ for each $\alpha$ and $j$

(c)$Q^{(\beta\alpha)}_{ji}=
\frac{{\overline Q^{(\alpha\beta)}_{ij}}}{\lambda^{(\alpha)}_i-
\lambda^{(\beta)}_j}$ (here the bar denotes complex conjugation).

(3) Finally, we  let $\Gamma$ be the union of $G$-orbits
\footnote{To be more precise, one has to take into account the action of 
an appropriate group of permutations on the set $\Delta$. This can be 
done as in Appendix 1 by working with a fundamental domain of $\Delta$ 
under this discrete action. } of elements of 
$\Delta$ under the action (\ref{Gaction}).
Gauge-fixing the action (\ref{ZADE})
gives the eigenvalue representation:
{\footnotesize \bea
\label{ADEev}
{\cal Z}=\int_{\gamma^{(1)}}{d\lambda^{(1)}_1}
\dots
\int_{\gamma^{(1)}}{d\lambda^{(1)}_{N_1}}
\dots 
\int_{\gamma^{(\rho)}}{d\lambda^{(\rho)}_1}
\dots
\int_{\gamma^{(\rho)}}{d\lambda^{(\rho)}_{N_\rho}}
\frac{\prod_{\alpha, i\neq j}{(\lambda^{(\alpha)}_i-
\lambda^{(\alpha)}_j)}}{\prod_{\alpha <\beta, i, j}{(\lambda^{(\alpha)}_i-
\lambda^{(\beta)}_j)^{s_{\alpha\beta}}}}
e^{-N\sum_{\alpha=1}^{\rho}{
\sum_{i=1}^{N_\alpha}W_\alpha(\lambda^{(\alpha)}_i)}}~~~~~~
\eea}
where we dropped constant prefactors and used the identity:
\be
\int_{\sigma}{du\wedge dv e^{-(\lambda-\lambda') u v}}=
-\frac{4\pi i}{\lambda-\lambda'}~~,
\ee
with $\lambda\neq \lambda'$ complex and 
$\sigma$ the contour in $\C^2$ given by:
\be
v=\frac{{\overline u}}{\lambda-\lambda'}~~.
\ee

Condition (\ref{reg}) acts as a `complex' regulator for the holomorphic model, 
by preventing common eigenvalues of neighboring 
$\Phi^{(\alpha)}$ and $\Phi^{(\beta)}$. 
As in the Hermitian case, working with the regularized model would 
therefore not suffice to recover the entire moduli space of planar solutions 
required by the conjecture of \cite{DV1}. 
To eliminate the constraints, the 
regularization condition (\ref{reg}) must be removed  
by taking the limit of coinciding $\gamma^{(\alpha)}$.
This limit can be performed in such a way that the end result is a
`renormalized' model which can be described as a statistical ensemble 
of `reduced' holomorphic models. We now show how this works for the 
case of holomorphic $A_2$ models. 

\subsection{Example: the holomorphic $A_2$ model}

For an $A_2$ quiver, one has $\rho=2$ and four matrices 
$\Phi^{(1)}$, $\Phi^{(2)}$, $Q^{(12)}$ and $Q^{(21)}$. 
The partition function
takes the form:
{\scriptsize \be
\label{A2ev}
{\cal Z}=\int_{\gamma^{(1)}}{d\lambda^{(1)}_1}
\dots
\int_{\gamma^{(1)}}{d\lambda^{(1)}_{N_1}} 
\int_{\gamma^{(2)}}{d\lambda^{(2)}_1}
\dots
\int_{\gamma^{(2)}}{d\lambda^{(2)}_{N_2}}
\frac{\prod_{i\neq j}{(\lambda^{(1)}_i-
\lambda^{(1)}_j)}\prod_{i\neq j}{(\lambda^{(2)}_i-
\lambda^{(2)}_j)}}{\prod_{i,j}{(\lambda^{(1)}_i-\lambda^{(2)}_j)}}
e^{-N\sum_{i=1}^{N_1}{W_1(\lambda^{(1)}_i)}-N\sum_{i=1}^{N_2}{W_2(\lambda^{(2)}_i)}}
\ee}
\noindent for two disjoint curves $\gamma^{(1)}$ and $\gamma^{(2)}$.

\subsubsection{Classical vacua}

Extremizing the action:
\be
S=N tr[W_1(\Phi^{(1)})+W_2(\Phi^{(2)})]+tr[Q^{(21)}\Phi^{(1)}Q^{(12)}-
Q^{(12)}\Phi^{(2)}Q^{(21)} ]
\ee
gives the equations:
\bea
\label{ceom}
NW_1'(\Phi^{(1)})&=&-Q^{(12)}Q^{(21)}~~,~~NW_2'(\Phi^{(2)})=Q^{(21)}Q^{(12)}~~,~~\nn\\
\Phi^{(2)}Q^{(21)}&=&Q^{(21)}\Phi^{(1)}~~~~,~~~~\Phi^{(1)}Q^{(12)}=Q^{(12)}\Phi^{(2)}~~.
\eea
Combining these, one easily obtains:
\bea
W'_1(\Phi^{(1)})[W'_1(\Phi^{(1)})+W'_2(\Phi^{(1)})]&=&0\nn\\ 
W'_2(\Phi^{(2)})[W'_1(\Phi^{(2)})+W'_2(\Phi^{(2)})]&=&0~~.
\eea
Assuming that $\Phi^{(1)}$ and $\Phi^{(2)}$ are diagonalizable with
eigenvalues $\lambda^{(1)}_i$ and $\lambda^{(2)}_j$, one finds:
\bea
\label{cv1}
W'_1(\lambda^{(1)}_i)[W'_1(\lambda^{(1)}_i)+W'_2(\lambda^{(1)}_i)]&=&0\nn\\ 
W'_2(\lambda^{(2)}_j)[W'_1(\lambda^{(2)}_j)+W'_2(\lambda^{(2)}_j)]&=&0~~.
\eea
On the other hand, the last row of equations in (\ref{ceom}) gives:
\bea
\label{cv2}
(\lambda^{(2)}_j-\lambda^{(1)}_i)Q^{(21)}_{ji}&=&0~~\nn\\
(\lambda^{(1)}_i-\lambda^{(2)}_j)Q^{(12)}_{ij}&=&0~~.
\eea
In the generic case when $W'_1$ and $W'_2$ have no common roots, 
equations (\ref{cv1}) and (\ref{cv2}) show that a typical classical vacuum is
specified (up to permutations of indices) by choosing:

(1) roots $\lambda^{(1)}_1\dots \lambda^{(1)}_{N_1-k}$ of $W'_1$, 

(2) roots $\lambda^{(2)}_1\dots \lambda^{(2)}_{N_2-k}$ of $W'_2$, 

(3) roots $\mu_1\dots \mu_k$ of $W'_1+W'_2$ such that
    $\lambda^{(1)}_{N-k+m}=\lambda^{(2)}_{N-k+m}=\mu_m$ for $m=1\dots k$. 

(4) some nonzero values for $Q^{(21)}_{N_2-k+m,N_1-k+m}$ and 
 $Q^{(12)}_{N_1-k+m,N_2-k+m}$ for $m=1\dots k$. In fact, one can set
$Q^{(12)}_{N_1-k+m,N_2-k+m}=1$ for all $m=1\dots k$ by using 
the gauge transformations.

Less generic vacua arise, for example, 
by allowing some of the $\mu_k$ to coincide. Notice
that vacua with $k\neq 0$ (i.e. vacua for which some $\lambda^{(1)}_i$
  coincide with some $\lambda^{(2)}_j$) are removed when imposing the condition
  $\gamma^{(1)}\cap \gamma^{(2)}=\emptyset$ --- such vacua will not be
  visible unless one removes this condition on the contours. 
We now proceed to remove this regulator, by taking the limit in which the two
curves coincide. 

\subsubsection{The limit of coinciding contours}

To study the limit when $\gamma^{(1)}$ and $\gamma^{(2)}$ coincide, we let 
$\gamma^{(2)}=\gamma$, $\gamma^{(1)}=\gamma+\eta n$ (where $n$ is the 
normal vector field to $\gamma$, chosen as in figure \ref{normal} of 
Section 3) and take the positive
quantity $\eta$ to zero (note that this requires $\gamma$ to asymptote to 
lines lying in the intersection of some convergence sectors of 
$W_1$ and $W_2$). 
For small $\eta$, we can use the length coordinate $s$ of 
$\gamma=\gamma^{(2)}$ 
also as a parameter along $\gamma^{(1)}$. 
We then have $\lambda^{(1)}(s)=\lambda(s)+\eta
n(s)$, where $\lambda(s)$ is the parameterization of $\gamma$.
For $\eta \rightarrow 0^+$, the Sokhotsky formulae
(\ref{Sokh}) give:
\be
\frac{1}{\lambda^{(1)}_i-\lambda^{(2)}_j}={\cal P}
\frac{1}{\lambda(s^{(1)}_i)-\lambda(s^{(2)}_j)}
-\frac{i\pi}{{\dot \lambda}(s^{(2)}_j)}\delta(s^{(1)}_i-s^{(2)}_j)~~.
\ee
Therefore, the denominator in the integrand of (\ref{A2ev}) takes the
following form:
{\footnotesize \bea
\label{S_expansion}
&&\lim_{\eta\rightarrow 0^+}
{\frac{1}{\prod_{i,j}{(\lambda_i^{(1)}-\lambda^{(2)}_j)}}}=\nn\\
&&\sum_{k=0}^{{\rm
    min}(N_1,N_2)}{\sum_{\tiny \begin{array}{c}1\leq i_1<\dots < i_k\leq N_1
      \\ 1\leq j_1<\dots < j_k\leq N_2\end{array}}}
\sum_{\sigma \in \Sigma_k}(-i\pi)^k
\frac{\delta(s^{(1)}_{i_1}-s^{(2)}_{j_{\sigma(1)}})
\dots \delta(s^{(1)}_{i_k}-s^{(2)}_{j_{\sigma(k)}})}
{{\dot \lambda}(s^{(2)}_{j_1})\dots {\dot \lambda}(s^{(2)}_{j_k})
}
\prod_{(i,j)\neq (i_1,j_{\sigma(1)})\dots (i_k,j_{\sigma(k)})}
{{\cal P}\frac{1}{\lambda^{(1)}_i-\lambda^{(2)}_j}}~~,~~~~~~~~~~~~~\nn\\
&+& ({\rm {\large ~higher~incidence~terms}})~~,
\eea}
where $\Sigma_k$ is the group of permutations on $k$ elements. 
The `higher incidence terms' are terms involving delta-function products 
of the type 
$\delta(s^{(1)}_i-s^{(2)}_j)\delta(s^{(1)}_i-s^{(2)}_k)=
\delta(s^{(1)}_i-s^{(2)}_j)\delta(s^{(2)}_j-s^{(2)}_k)$ with distinct
$i,j,k$. Such terms can be neglected since --- as we shall see in a moment ---
they do not contribute to the final result. 

Substituting (\ref{S_expansion}) in (\ref{A2ev}) gives:
\be
\label{ens}
{\cal Z}_{lim}(\gamma)=
\sum_{k=0}^{{\rm min}(N_1,N_2)}{(-i\pi)^k \frac{N_1!N_2!}{k! (N_1-k)!(N_2-k)!}
  {\cal Z}_k(\gamma)}~~,
\ee
where:
 \bea
{\cal Z}_k(\gamma)&=&
\int_{\gamma}{d\mu_1}\dots \int_{\gamma}{d\mu_k}
\int_{\gamma}{d\lambda^{(1)}_1} \dots
  \int_{\gamma}{d\lambda^{(1)}_{N_1-k}}\int_{\gamma}{d\lambda^{(2)}_1} \dots
  \int_{\gamma}{d\lambda^{(2)}_{N_2-k}}~\Delta_k~e^{-N S_k}~~,~~~~~~~
\eea
with:
 \be
\Delta_k:={\tilde \Delta}_k{\cal P}\frac{1}
{\prod_{\tiny \begin{array}{c}i=1\dots N_1-k\\j=1\dots N_2-k\end{array}}
{(\lambda^{(1)}_i-\lambda^{(2)}_j)}}~~,
\ee
\bea
{\tilde \Delta}_k&=&
{\prod}'_{\tiny i,j=1\dots N_1-k}{(\lambda^{(1)}_i- \lambda^{(1)}_j)}
{\prod}'_{\tiny i,j=1\dots N_2-k}{(\lambda^{(2)}_i- \lambda^{(2)}_j)}
{\prod}'_{\tiny i,j=1\dots k}{(\mu_i- \mu_j)}\nn\\
& & \prod_{\tiny \begin{array}{c}i=1\dots N_1-k\\ j=1\dots k
\end{array}}{(\mu_j-\lambda^{(1)}_i)}
\prod_{\tiny \begin{array}{c}i=1\dots N_2-k\\ j=1\dots k
\end{array}}{(\lambda^{(2)}_i-\mu_j)}~~
\eea
and: 
\be
S_k:=\sum_{j=1}^{N_1-k}{W_1(\lambda^{(1)}_j)}+
\sum_{j=1}^{N_2-k}{W_1(\lambda^{(2)}_j)}+
\sum_{j=1}^k{\left[W_1(\mu_j)+W_2(\mu_j)\right]}~~.
\ee

Notice that the `higher incidence terms' of equation (\ref{S_expansion}) 
bring zero contribution to (\ref{ens}). This is due to the extra-factors 
of $\mu_j-\mu_k$ contributed by the two products in the numerator of 
the weight factor of (\ref{A2ev}).

{\bf Observation} 
Consider a quantity $H(z)$ (which depends on $\lambda^{(\alpha)}_j$ 
but is symmetric under separate permutations of $\lambda^{(1)}_1\dots 
\lambda^{(1)}_{N_1}$ and of $\lambda^{(2)}_1\dots \lambda^{(2)}_{N_2}$). 
Then its average $\langle H(z)\rangle$ has the following behavior in the limit 
$\eta\rightarrow 0^+$:
\be
\lim_{\eta \rightarrow 0}{\langle H(z)\rangle}=
\frac{\sum_{k=0}^{{\rm min}(N_1,N_2)}{
(-i\pi)^k \frac{N_1!N_2!}{k! (N_1-k)!(N_2-k)!}  
{\cal Z}_k \langle H(z)\rangle_k
}}{\sum_{k=0}^{{\rm min}(N_1,N_2)}{(-i\pi)^k 
\frac{N_1!N_2!}{k! (N_1-k)!(N_2-k)!}
  {\cal Z}_k(\gamma)}}
:=\langle H(z) \rangle_{lim}~~,
\ee
where:
{\footnotesize \be
\langle H(z)\rangle_k=\frac{1}{{\cal Z}_k}
\int_{\gamma}{d\mu_1}\dots \int_{\gamma}{d\mu_k}
\int_{\gamma}{d\lambda^{(1)}_1} \dots
  \int_{\gamma}{d\lambda^{(1)}_{N_1-k}}\int_{\gamma}{d\lambda^{(2)}_1} \dots
  \int_{\gamma}{d\lambda^{(2)}_{N_2-k}}
\Delta_k H(z)e^{-N S_k}~~.
\ee}
Thus $\langle H(z)\rangle_{lim}$ is simply a weighted average taken over the
limiting ensemble.

\subsubsection{Equations of motion for the limiting ensemble}

Let us fix a component $k$ of the limiting ensemble, and work with the
model defined by the partition function ${\cal Z}_k$.
Writing $\Delta_k$ as an exponential gives:
\be
{\cal Z}_k(\gamma)=\int{d\sigma_1}\dots \int{d\sigma_k}
\int{ds^{(1)}_1} \dots
  \int{ds^{(1)}_{N_1-k}}\int{ds^{(2)}_1} \dots
  \int{ds^{(2)}_{N_2-k}}~e^{-N S^{eff}_k}~~,~~~~~~~
\ee
where:
{\footnotesize \bea
S_k^{eff} &=& S_k
-\frac{1}{N}
{\sum}'_{\tiny i,j=1\dots N_1-k}{\ln (\lambda^{(1)}_i- \lambda^{(1)}_j)}
-\frac{1}{N}
{\sum}'_{\tiny i,j=1\dots N_2-k}{\ln (\lambda^{(2)}_i- \lambda^{(2)}_j)}
-\frac{1}{N}{\sum}'_{\tiny i,j=1\dots k}{\ln (\mu_i- \mu_j)}\nn\\
&-& \frac{1}{N}\sum_{\tiny \begin{array}{c}i=1\dots N_1-k\\ j=1\dots k
\end{array}}{\ln (\mu_j-\lambda^{(1)}_i)}
-\frac{1}{N}\sum_{\tiny \begin{array}{c}i=1\dots N_2-k\\ j=1\dots k
\end{array}}{\ln (\lambda^{(2)}_i-\mu_j)}
+\frac{1}{N}
\sum_{\tiny \begin{array}{c}i=1\dots N_1-k\\j=1\dots N_2-k\end{array}}
{\ln (\lambda^{(1)}_i-\lambda^{(2)}_j)}~~\nn\\
&-&\frac{1}{N}\sum_{j=1}^{N_1-k}{\ln {\dot \lambda}(s^{(1)}_j)}
-\frac{1}{N}\sum_{j=1}^{N_2-k}{\ln {\dot \lambda}(s^{(2)}_j)}
-\frac{1}{N}\sum_{j=1}^{k}{\ln {\dot \lambda}(\sigma_j)}~~~~~~~~~~~~~~~~~
\eea}
and we wrote $\lambda^{(1)}_i=\lambda(s^{(1)}_i),
\lambda^{(2)}_i=\lambda(s^{(2)}_i)$ and $\mu_i=\lambda(\sigma_i)$. 
Extremizing this with respect to $s^{(\alpha)}_i$ and $\sigma_i$ gives the 
equations of motion:
{\footnotesize \bea
\label{A2eom}
\frac{2}{N}\sum_{j=1}^{N_1-k~,}{\frac{1}{\lambda(s^{(1)}_i)-\lambda(s^{(1)}_j)}}
-\frac{1}{N}\sum_{j=1}^{N_2-k}{\frac{1}{\lambda(s^{(1)}_i)-\lambda(s^{(2)}_j)}}
&+&\frac{1}{N}\sum_{j=1}^{k}{\frac{1}{\lambda(s^{(1)}_i)-\lambda(\sigma_j)}}=
\nn\\
&=&W'_1(\lambda(s^{(1)}_i))-
\frac{1}{N}\frac{{\ddot \lambda}(s_i^{(1)})}{{\dot \lambda}(s^{(1)}_i)^2}
~~~~~~~~~~~~~~~~~~~~~\nn\\
\frac{2}{N}\sum_{j=1}^{N_2-k~,}{\frac{1}{\lambda(s^{(2)}_i)-\lambda(s^{(2)}_j)}}
-\frac{1}{N}\sum_{j=1}^{N_1-k}{\frac{1}{\lambda(s^{(2)}_i)-\lambda(s^{(1)}_j)}}
&+&\frac{1}{N}\sum_{j=1}^{k}{\frac{1}{\lambda(s^{(2)}_i)-\lambda(\sigma_j)}}=
~~~~~~~~~~~~\\
&=&W'_2(\lambda(s^{(2)}_i))-
\frac{1}{N}\frac{{\ddot \lambda}(s_i^{(2)})}{{\dot \lambda}(s^{(2)}_i)^2}
~~~~~~~~~~~~~~~~~~~~~~~~\nn\\
\frac{2}{N}\sum_{i,j=1}^k{\frac{1}{\lambda(\sigma_i)-\lambda(\sigma_j)}}+
\frac{1}{N}\sum_{j=1}^{N_1-k~,}{\frac{1}{\lambda(\sigma_i)-\lambda(s^{(1)}_j)}}
+\frac{1}{N}\sum_{j=1}^{N_2-k}{\frac{1}{\lambda(\sigma_i)-\lambda(s^{(2)}_j)}}
\nn\\
&=&W'_1(\lambda(\sigma_i))+W'_2(\lambda(\sigma_i))-
\frac{1}{N}\frac{{\ddot \lambda}(\sigma_i)}{{\dot \lambda}(\sigma_i)^2}~~
~~~~~~~~~~~~~~~~~~~\nn
\eea}
Let us introduce the spectral densities:
\be
\rho^{(\alpha)}(s)
:=\frac{1}{N}\sum_{j=1}^{N_\alpha-k}{\delta(s-s^{(\alpha)}_j)}+
\frac{1}{N}\sum_{j=1}^{k}{\delta(s-\sigma_j)}~~,
\ee
with the normalization:
\be
\label{A2norm}
\int{ds \rho^{(\alpha)}(s)}=\frac{N_\alpha}{N}~~.
\ee
We also introduce the traced resolvents:
\be
\omega^{(\alpha)}(z):=\frac{1}{N}\sum_{j=1}^{N_\alpha-k}{\frac{1}{z-
\lambda(s^{(\alpha)}_j)}}+
\frac{1}{N}\sum_{j=1}^{k}{\frac{1}{z-
\lambda(\sigma_j)}}
=\int{ds\frac{\rho^{(\alpha)}(s)}{
z-\lambda(s)}}~~.
\ee

\subsubsection{The large $N$ Riemann surface}
For every quantity $\phi$, we define 
the large $N$ expansion coefficients $\phi_j$ as in equations (\ref{phi_exp}).
In particular, we have the planar limits $\rho^{(\alpha)}_{0}$
and:
\be
\label{omega_lim}
\omega^{(\alpha)}_{0}(z)=
\int{ds \frac{\rho^{(\alpha)}_{0}(s)}{z-\lambda(s) }}~~.
\ee
In the large $N$ limit, the eigenvalues $\lambda(s^{(1)}_i)$,
$\lambda(s^{(2)}_j)$ and $\lambda(\sigma_k)$ accumulate on curve segments 
sitting along $\gamma$. The planar limits (\ref{omega_lim})
will have cuts along three types of curvilinear intervals:

\

(1) loci $C_{13}^a$ resulting from a planar 
distribution of the eigenvalues $\lambda(s^{(1)}_j)$ 
with $j=1\dots N_1-k$; these will be cuts of $\omega^{(1)}_0$~~

(2) loci $C_{23}^b$ supporting a distribution 
of the eigenvalues $\lambda(s^{(2)}_j)$ with $j=1\dots N_2-k$; these 
give cuts of $\omega^{(2)}_0$

(3) loci $C_{12}^c$ resulting from a
distribution of $\lambda(\sigma_j)$ with $j=1\dots k$; they give 
common cuts of $\omega^{(1)}_0$ and $\omega^{(2)}_0$ . 

\

Here $a,b,c$ are indices counting the various occurrences of each type of cut. 
Note that the third type of locus can only arise from a component of the 
limiting ensemble for which $\frac{k}{N}$ has a finite limit as 
$N\rightarrow \infty$. Thus cuts of type $C_{12}$ can only develop in the 
large $N$ limit of the `renormalized' model with $\gamma^{(1)}=\gamma^{(2)}$.

The planar limit of the equations of motion (\ref{A2eom}) gives:
\bea
\label{A2eom0}
2\int{ds'}{\frac{\rho_0^{(1)}(s')}{\lambda(s)-\lambda(s')}}-
\int{ds'}{\frac{\rho_0^{(2)}(s')}{\lambda(s)-\lambda(s')}}&=&
W_1'(\lambda(s))~~,~~\lambda(s)\in C_{13}^a
\nn\\
2\int{ds'}{\frac{\rho_0^{(2)}(s')}{\lambda(s)-\lambda(s')}}-
\int{ds'}{\frac{\rho_0^{(1)}(s')}{\lambda(s)-\lambda(s')}}&=&W_2'(\lambda(s))
~~,~~\lambda(s)\in C_{23}^b\\
\int{ds'}{\frac{\rho_0^{(1)}(s')}{\lambda(s)-\lambda(s')}}+
\int{ds'}{\frac{\rho_0^{(2)}(s')}{\lambda(s)-\lambda(s')}}&=&W_1'(\lambda(s))+
W_2'(\lambda(s))~~,~~\lambda(s)\in C_{12}^c~~.\nn
\eea
These relations act as large $N$ saddle point equations 
for the limiting ensemble (\ref{ens}).
They also represent the `quantum' version of 
the three branches $W'_1(\lambda)=0$, $W'_2(\lambda)=0$ 
and $W'_1(\lambda)+W'_2(\lambda)=0$ 
of the classical equations of motion (\ref{cv1}). 
It is also clear that we have:
\be
\label{rho_sym}
\rho^{(1)}_{0}(s)=\rho^{(2)}_{0}(s)~~,~~\lambda(s)\in C^c_{12}~~.
\ee
This is the analogue of the classical relations 
$\lambda^{(1)}_i=\lambda^{(2)}_i$
on the corresponding branch of the moduli space (see point (3) of Subsection
5.1.1). 

As explained in Appendix 3, one can use (\ref{A2eom0}) 
and certain partial fraction decompositions
to derive the algebraic constraints:
{
\bea
\label{leqs0}
\omega^{(1)}_{0}(z)^2-\omega^{(1)}_{0}(z)
\omega^{(2)}_{0}(z)+\omega^{(2)}_{0}(z)^2-W_1'(z)
\omega^{(1)}_{0}(z)
-W_2'(z)\omega^{(2)}_{0}(z)+f^{(1)}_{0}(z)+f^{(2)}_{0}(z)
&=&0~~\nn\\
\omega^{(1)}_{0}(z)^2\omega^{(2)}_{0}(z)-
W'_1(z)
\left[\omega^{(1)}_{0}(z)^2+f_0^{(1)}(z)-W'_1(z)\omega^{(1)}_{0}(z)
\right]+g^{(1)}_0(z)- (1\leftrightarrow 2)&=&0~~~~~~~~~~
\eea}
where $f^{(\alpha)}_0, g^{(\alpha)}_0$ are polynomials defined through:
\be
\label{f_planar}
f^{(\alpha)}_{0}(z):=\int{ds
  \rho^{(\alpha)}_{0}(s)\frac{W_\alpha'(z)-W_\alpha'
(\lambda(s))}{z-\lambda(s)}}~~
\ee
and
\bea
\label{g_p}
g^{(1)}_{0}(z)&:=&
\int{d\alpha}\int{d\beta\left[W'_1(z)-W'_1(\lambda(\alpha))\right]
\frac{\rho^{(1)}_{0}(\alpha)\rho^{(2)}_{0}(\beta)}
{(\lambda(\alpha)-\lambda(\beta))
(z-\lambda(\alpha))}}~~\nn\\
g^{(2)}_{0}(z)&:=&
\int{d\alpha}\int{d\beta\left[W'_2(z)-W'_2(\lambda(\alpha))\right]
\frac{\rho^{(2)}_{0}(\alpha)\rho^{(1)}_{0}(\beta)}
{(\lambda(\alpha)-\lambda(\beta))
(z-\lambda(\alpha))}}~~.
\eea
Note that the normalization conditions (\ref{A2norm}) imply:
\be
\label{flead}
{\rm lcoeff}(f_0^{(\alpha)})=\frac{N_\alpha}{N}
{\rm deg} (W_\alpha) {\rm lcoeff}(W_\alpha)~~~.
\ee
where ${\rm lcoeff}(...)$ denotes the leading coefficient. 
Writing:
\bea
\label{A2shift}
\omega^{(1)}_{0}(z)&=&u_1(z)+t_1(z)\nn\\
\omega^{(2)}_{0}(z)&=&-u_2(z)+t_2(z)~~,
\eea
where:
\bea
\label{tdef}
t_1(z)&:=&\frac{2W'_1(z)+W'_2(z)}{3}~~\nn\\
t_2(z)&:=&\frac{2W'_2(z)+W'_1(z)}{3}
\eea
brings the constraints (\ref{leqs0})  to the form:
\bea
\label{Viete}
u_1(z)^2+u_1(z)u_2(z)+u_2(z)^2&=&p(z)~~\nn\\
u_1(z)^2u_2(z)+u_1(z)u_2(z)^2~~~&=&-q(z)~~,
\eea
where:
\bea
\label{pq_def}
p(z)&:=&t_1(z)^2-t_1(z)t_2(z)+t_2(z)^2-f^{(1)}_0(z)-f^{(2)}_0(z)~~~\\
q(z)&:=&-t_1(z)t_2(z)\left[t_1(z)-t_2(z)\right]-t_1(z)f^{(2)}_0(z)+
t_2(z)f^{(1)}_0(z)-g_0(z)~~.\nn
\eea
In the last equation, we introduced the 
polynomial $g_0(z)=g^{(1)}_0(z)-g^{(2)}_0(z)$ which has the 
following explicit form in terms of matrix model data:
\bea
\label{g_planar}
g_0(z)&=&\int{d\alpha}\int{d\beta\left[W'_1(z)-W'_1(\lambda(\alpha))\right]
\frac{\rho^{(1)}_{0}(\alpha)\rho^{(2)}_{0}(\beta)}
{(\lambda(\alpha)-\lambda(\beta))
(z-\lambda(\alpha))}}-\nn\\
&-&\int{d\alpha}\int{d\beta\left[W'_2(z)-W'_2(\lambda(\alpha))\right]
\frac{\rho^{(2)}_{0}(\alpha)\rho^{(1)}_{0}(\beta)}
{(\lambda(\alpha)-\lambda(\beta))
(z-\lambda(\alpha))}}~~
\eea

Defining $u_3(z):=-u_1(z)-u_2(z)$, identities
(\ref{Viete}) can be recognized as the Viete relations 
of the cubic:
\be
\label{A2curve}
u^3-p(z)u-q(z)=0~~
\ee
when the left hand side is viewed as a polynomial in $u$. 
This shows that $u_1(z)$, $u_2(z)$ and $u_3(z)$ are the three branches of the
affine algebraic curve (\ref{A2curve}).  This is the precise form of the curve
suggested in \cite{DV3} (where the explicit form of the polynomials $p,q$ in
terms of matrix model data was not given). Note that the left hand side 
of (\ref{A2curve}) can also be written as:
\be
\label{pert_curve}
(u-u_1)(u-u_2)(u-u_3)=(u+t_1)(u-t_2)(u-t_1+t_2)+
(f^{(1)}_0+f^{(2)}_0)u+t_1f^{(2)}_0-t_2
f^{(1)}_0+g_0~~.
\ee

Studying the behavior of $u_j$ near the branch cuts of (\ref{A2curve}) 
allows one to identify these as the loci $C_{13}^a$, $C_{23}^b$ and $C_{12}^c$
where the eigenvalues accumulate; then the jump equations 
across these cuts can be seen to be 
equivalent with the planar equations of motion (\ref{A2eom0}). 
Below, we shall use this reasoning in order to give a reconstruction 
theorem for the holomorphic $A_2$ model, similar to the one we found 
in Section 3 for the holomorphic one-matrix model.

\subsubsection{The reconstruction theorem}

Let us start with a curve of form (\ref{A2curve}) with $p,q$ given by 
(\ref{pq_def}), where 
$f^{(\alpha)}_0$ and $g_0$ are complex polynomials of
degree $n-2$ subject to the constraints (\ref{flead}).
Given such data, we show how one can construct a curve $\gamma$ and 
distributions $\rho^{(\alpha)}_0(s)$ 
along this curve such that relations (\ref{A2norm}), 
(\ref{omega_lim}), (\ref{A2eom0}), (\ref{rho_sym}) 
and (\ref{f_planar}), (\ref{g_planar}) hold.  

Using expression (\ref{pert_curve}), one finds the 
following asymptotic behavior for large $z$:
\bea
\label{u_as}
u_1(z)&=&-t_1+\frac{{\rm lcoeff}(f^{(1)}_0)}{deg (W_1) {\rm lcoeff}(W_1)}
+O(1/z^2)~~\nn\\
u_2(z)&=&+t_2+\frac{{\rm lcoeff}(f^{(2)}_0)}{deg (W_2) {\rm lcoeff}(W_2)}
+O(1/z^2)~~.
\eea
In particular, we can use these asymptotic forms in order 
to fix the indexing of 
branches for (\ref{curve}) (i.e. $u_1$ is the branch which asymptotes to 
$-t_1$, while $u_2$ is the branch which asymptotes to $+t_2$). 

When viewed as a triple cover of the $z$-plane, the curve (\ref{A2curve}) has
three types of cuts, namely those connecting the pairs of branches $(1,3)$,
$(2,3)$ and $(1,2)$. Denote these cuts by $C_{13}^a,C_{23}^b$ and $C_{12}^c$.
Picking $\gamma$ to contain all cuts, we let $\omega^{(1)}_0$ and 
$\omega_0^{(2)}$ be given by relations (\ref{A2shift}) and define:
{\bea
\label{A2rho_def}
\rho_0^{(1)}(s)&:=&
\frac{1}{2\pi i}{\dot \lambda}(s)
\left[\omega^{(1)}_0(\lambda(s)-i0)-\omega^{(1)}_0
(\lambda(s)+i0)
\right]~~,~~{\rm for}~~ \lambda(s)\in C_{13}^a~~~~~~\nn\\
\rho_0^{(2)}(s)&:=&
\frac{1}{2\pi i}{\dot \lambda}(s)
\left[\omega^{(2)}_0(\lambda(s)-i0)-\omega^{(2)}_0
(\lambda(s)+i0)
\right]~~,~~{\rm for}~~ \lambda(s)\in C_{23}^b~~,\nn\\
\rho^{(1)}(s)=\rho_0^{(2)}(s)&:=&
\frac{1}{2\pi i}{\dot \lambda}(s)
\left[\omega^{(1)}_0(\lambda(s)-i0)-\omega^{(1)}_0
(\lambda(s)+i0)
\right]\\
&=&\frac{1}{2\pi i}{\dot \lambda}(s)
\left[\omega^{(2)}_0(\lambda(s)-i0)-\omega^{(2)}_0
(\lambda(s)+i0)
\right]~~,~~{\rm for}~~ \lambda(s)\in C_{12}^c~~.\nn
\eea}
The very last equality in these relations follows by using (\ref{A2shift}) and 
the fact that $u_3=-u_1-u_2$ is continuous across cuts of type $C_{12}$.  
In particular, this means that (\ref{rho_sym}) holds.
Extending $\rho_0^{(\alpha)}$ by zero to the entire curve $\gamma$, 
relations (\ref{A2rho_def}) and the Sokhotsky formulae 
immediately imply that equations (\ref{omega_lim}) hold. In turn, these 
relations, the asymptotic behavior (\ref{u_as}) and 
the constraints (\ref{flead}) on the leading coefficients 
imply the normalization conditions (\ref{A2norm}). 

We next notice that:
\bea
\label{uij}
u_1(z)-u_3(z)&=&2u_1(z)+u_2(z)=2\omega^{(1)}_0(z)-\omega^{(2)}_0
(z)-W'_1(z)\nn\\
u_2(z)-u_3(z)&=&u_1(z)+2u_2(z)=\omega^{(1)}_0(z)-2\omega^{(2)}_0(z)+W'_2(z)\\
u_1(z)-u_2(z)&=&\omega^{(1)}_0(z)+\omega^{(2)}_0(z)-(W'_1(z)+W'_2(z))~~,\nn
\eea
where we used relations (\ref{A2shift}) and (\ref{tdef}).
Since a cut of type 
$C_{\alpha\beta}$ connects the branches $u_\alpha$ and $u_\beta$, 
we have the jump equations:
\bea
\label{rho_omega}
& &u_\alpha(\lambda \pm 0n(\lambda))
=u_\beta(\lambda\mp i 0n(\lambda))\Rightarrow\\ 
& &\left[u_\alpha(\lambda +0n(\lambda))-u_\beta(\lambda + 0n(\lambda))
\right]+
\left[u_\alpha(\lambda -0n(\lambda))-u_\beta(\lambda - 0n(\lambda))\right]
=0~~\nn
\eea
along such a cut. Combining this with relations 
(\ref{uij})
and using equations (\ref{omega_lim}) and the Sokhotsky formulae 
(\ref{Sokh}) shows that 
$\rho^{(\alpha)}_0$ satisfy the planar equations of motion (\ref{A2eom0}). 

To prove that  (\ref{f_planar}) and (\ref{g_planar}) 
hold, one can now simply repeat the 
original argument (found in Appendix 3) 
leading to the large $N$ curve (\ref{A2curve}); this is 
possible since we already showed that all assumptions of that argument 
(namely relations  
(\ref{omega_lim}), (\ref{A2eom0}), (\ref{rho_sym})) hold. This shows 
that equations (\ref{Viete}) and (\ref{pq_def}) 
must hold with polynomials $f,g$ defined 
by relations (\ref{f_planar}) and (\ref{g_planar}). 
Since by hypothesis the same relations hold 
with the original polynomials $f^{(\alpha)}_0$ and $g_0$, it 
immediately follows that the latter can indeed be expressed in the form
(\ref{f_planar}) and (\ref{g_planar}). This concludes the proof that the 
`renormalized' holomorphic $A_2$ model indeed probes the 
whole moduli space of (\ref{A2curve}) with the constraints (\ref{flead}).

\paragraph{Observation} 

The third `branch' of the planar equations of motion 
(\ref{A2eom0}) is only allowed in the 
`renormalized' model described by the limiting ensemble (\ref{ens}). 
If one works with the regularized model instead (the original 
model for which $\gamma^{(1)}$ and $\gamma^{(2)}$ are disjoint), then cuts of 
type $C^c_{12}$ are not allowed. Indeed, such cuts connect the branches 
$u_1$ and $u_2$, and thus are cuts for both $\omega_0^{(1)}$ and 
$\omega_0^{(2)}$, 
which would require $C_{12}^c\subset \gamma^{(1)}\cap \gamma^{(2)}$, in 
contradiction with the regularization condition 
$\gamma^{(1)}\cap \gamma^{(2)}=\emptyset$. 
Hence the regularized model can only probe that part of the
moduli space of (\ref{A2curve}) for which all cuts of type $C_{12}^c$ are 
reduced to double points. 
This is similar to what happens for the Hermitian $A_2$ model, as discussed at
the beginning of this section.  In particular, the regularized
representation (\ref{A2ev}) cannot capture the entire family of curves
(\ref{A2curve}), and, in fact, can only describe a subvariety in the space of 
all allowed polynomials $f^{(\alpha)}_0$ and $g^{(\alpha)}_0$. 
To explore the full moduli space, one {\em must} consider the limiting
ensemble (\ref{ens}). In the context of the Dijkgraaf-Vafa conjecture, 
the dual field theory is an $N=1$ supersymmetric field theory derived from an 
$A_2$ quiver (such theories have been studied in \cite{Cachazo, Radu}). 
This field 
theory is certainly allowed to explore the branch whose classical description 
is given by $W'_1(\lambda)+W'_2(\lambda)=0$. Therefore,  
consideration of the limiting ensemble (\ref{ens}) of the holomorphic model 
is required by the conjecture of \cite{DV1}
as applied to the $A_2$ quiver field theories of 
\cite{Cachazo}. A similar analysis can be performed for general 
$ADE$ models, with the same conclusion. 

\acknowledgments{I thank A. Klemm for support and K. Landsteiner for
  many useful conversations.  This work was supported by DFG project 
KL1070/2-1.}

\appendix

\section{Integration over gauge group orbits for the holomorphic one-matrix 
model}

In this Appendix, we 
show how the eigenvalue representation (\ref{ev}) results from 
the matrix integral (\ref{hmm}).

\subsection{Orbit decomposition of ${\cal M}$}

Let $\Delta$ be the set of diagonal $N\times N$ complex 
matrices $D$ with distinct eigenvalues:
\be
\Delta:=\{D={\rm diag}(\lambda_1\dots \lambda_N)|\lambda_j\in \C \rm{~and~} 
\prod_{i\neq j}{(\lambda_i-\lambda_j)}\neq 0\}~~
\ee
and fix a fundamental domain $\Delta_0$ 
for the obvious action of the permutation group $\Sigma_N$ on $\Delta$.
Then relation (\ref{diag}) gives the orbit decomposition:
\be
\label{od0}
{\cal M}=\sqcup_{D\in \Delta_0}{{\cal O}_D}~~,
\ee
where ${\cal O}_D$ is the $GL(N,\C)$-orbit 
of a matrix $D\in \Delta_0$ under the similarity action 
(\ref{sim}). 

The stabilizer of each $D\in \Delta_0$ in $GL(N,\C)$ is the subgroup  
$T_N\approx (\C^*)^N$ consisting of diagonal matrices.
Hence each orbit has the form:
\be
{\cal O}_D=H:=GL(N,\C)/T_N~~,
\ee
where the homogeneous space $H$ has dimension $N^2-N$ (here $T_N$ acts 
on $GL(N,\C)$ from the right, i.e. $S\rightarrow ST$ for $S\in GL(N,\C)$ and
$T\in T_N$). The 
orbit decomposition (\ref{od0}) takes the form:
\be
{\cal M}=H\times \Delta_0~~.
\ee

\subsection{Decomposition of $w$}

Let:
\be
w_\Delta=\prod_{j=1}^N{d\lambda_j}
\ee
be the translation-invariant holomorphic volume form on $\Delta$ and:
\be
w_H=\wedge_{i\neq j}{\omega_{ij}}
\ee
be the left-invariant holomorphic volume form on the complex 
homogeneous space $H$, where $\omega=S^{-1}dS$ is the matrix whose elements
give a basis of left-invariant holomorphic one-forms on $GL(N,\C)$. 

If $l_S(S'):=SS'$ is the left action of $GL(N,\C)$ on $H$, then $w_H$ 
satisfies:
\be
\label{li}
(l_S)^*w_H=w_H~~,~~S\in GL(N,\C)~~.
\ee
Using the projections $\pi_H$ and $\pi_\Delta$ of 
${\cal M}=H\times \Delta_0$ 
onto its two factors, we define a holomorphic $N^2$-form on 
${\cal M}$ by:
\be
\label{w_0}
w_0:=\pi_H^*(w_H)\wedge \pi_\Delta^*(w_\Delta)~~.
\ee
For dimension reasons, this must be related to $w$ through:
\be
\label{decomp}
w=f w_0~~,
\ee
where $f(M)$ is a holomorphic function on ${\cal M}$. 
Using the left-invariance 
(\ref{li}) of $w_H$, it is easy to check that $w_0$ is invariant
\footnote{To check this, notice that $\pi_H\circ \tau(S)=l_S\circ \pi_H$.} 
under the action (\ref{sim}):
\be
\tau(S)^*(w_0)=w_0~~.
\ee 
Using $GL(N,\C)$ invariance of $w$ and $w_0$, relation 
(\ref{decomp}) implies that $f$ is invariant:
\be
\label{finvar}
f(SMS^{-1})=f(M)~~.
\ee
In particular, we have $f(M)=f(D)$ if $SMS^{-1}=D$ with $D$ diagonal. 
It thus suffices to determine the values of $f$ for diagonal matrices $D$. 

For this, we first describe the the cotangent space to ${\cal M}$ at the 
points of $\Delta_0$ (where $\Delta_0$ is viewed as a subset of ${\cal M}$). 
In the vicinity of $\Delta_0$, we can write 
$M=SDS^{-1}\approx (1+A)D(1-A)\approx D+[A,D]$, 
where $S=e^A\approx 1+A$ with $A$ an infinitesimal 
generator of $GL(N,\C)$. 
Therefore, at a point $D$ of $\Delta_0$, we have:
\be
\label{dM}
dM=dD+[dA,D]\Longleftrightarrow dM_{ij}=\delta_{ij}d\lambda_i+
(\lambda_j-\lambda_i)dA_{ij}~~.
\ee
Note that there is no $[A,dD]$ piece in the right hand side of this relation,
since we compute the form $dM$ at the point $D$ (where $A=0$). 

Noticing that $w_H=\wedge_{i\neq j}{dA_{ij}}$ at such a point, 
relation (\ref{dM}) gives the form of $w$ at $M=D$:
\be
w= (-1)^{N^2(N-1)/2}\wedge_{i,j}{dM_{ij}}=
(-1)^{N^2(N-1)/2} \left[\prod_{i\neq j}{(\lambda_i-\lambda_j)}\right] w_0~~.
\ee
Comparing with (\ref{decomp}), we find that 
$f$ is given by the usual Vandermonde determinant: 
\be
f(M)=f(D)=\prod_{i\neq j}{(\lambda_i-\lambda_j)}~~.
\ee
Combining this with (\ref{decomp}) gives the following expression for 
the holomorphic measure:
\be
\label{w}
w=(-1)^{N^2(N-1)/2} \prod_{i\neq j}{(\lambda_i-\lambda_j)}w_0~~,
\ee
where $\lambda_i$ are the eigenvalues of the matrix $M$ at which we evaluate 
$w$. 

\subsection{The eigenvalue representation}

The $GL(N,\C)$ invariance of the action, relation (\ref{w})
and the decomposition (\ref{w_0}) allow
us to perform the integral over the gauge-group variables in the partition
function (\ref{hmm}):
\be
\label{ev_intmd}
{\tilde {\cal Z}}_N(\gamma,t)=\frac{1}{{\cal N}}(-1)^{N^2(N-1)/2}
hvol(H) \int_{\Delta_0(\gamma)}{\prod_{j=1}^N{d\lambda_j}
\prod_{i\neq j}{(\lambda_i-\lambda_j)} e^{-N\sum_{j=1}^N{W(\lambda_j)}}}
\ee 
where $hvol(H)=\int_H{\omega_H}$ is the  
`holomorphic volume' of $H$ and:
\be
\Delta_0(\gamma)=\{D={\rm diag}(\lambda_1\dots \lambda_N)\in \Delta_0
|\lambda_j\in \gamma~,~ \forall j=1\dots N \}~~.
\ee
Since this relation holds for {\em any} choice of fundamental domain 
$\Delta_0$, we can write (\ref{ev_intmd}) in the form:
\be
{\tilde {\cal Z}}_N(\gamma,t)=\frac{1}{{\cal N}}(-1)^{N^2(N-1)/2}
\frac{1}{N !}~hvol(H) {\cal Z}_N(\gamma,t)~~,
\ee 
where:
\be
\label{ev0}
{\cal Z}_N(\gamma,t)=
\int_{\Delta(\gamma)}{\prod_{j=1}^N{d\lambda_j}
\prod_{i\neq j}{(\lambda_i-\lambda_j)} e^{-N\sum_{j=1}^N{W(\lambda_j)}}}
\ee
with:
\be
\label{Delta_gamma}
\Delta(\gamma)=\{D={\rm diag}(\lambda_1\dots \lambda_N)|\lambda_j\in \gamma~,~ 
\forall j=1\dots N \rm{~and~} \prod_{i\neq j}{(\lambda_i-\lambda_j)}\neq 0
\}~~
\ee
This gives the eigenvalue representation (\ref{ev}). 

{\bf Observation} When writing the representation (\ref{ev}), we have tacitly 
extended the integral to allow for colliding eigenvalues
$\lambda_i=\lambda_j$ (this is certainly allowed, since the integrand of 
(\ref{ev0}) is 
well-behaved there). This amounts to 
treating non-diagonalizable matrices 
by a point-splitting prescription, and can be formulated 
in more detail by working with:
\be
\Delta_\epsilon:=
\{D={\rm diag}(\lambda_1\dots \lambda_N)|\lambda_j\in \C \rm{~and~} 
|\lambda_i-\lambda_j|>\epsilon ~~{\rm for}~~i\neq j\}~~
\ee 
instead of $\Delta$ and with a similar modification ${\cal M}_\epsilon$ 
of ${\cal M}$. Then one defines the partition function as the limit 
$\epsilon\rightarrow 0^+$ of the regularized partition function obtained 
in this manner. It is easy to adapt the derivation given above in order to 
include explicitly such a regulator. The result, however, is the same 
as (\ref{ev}), because the integrand of (\ref{ev0}) is well-behaved 
when eigenvalues come close to each other.

\section{Example of the relevance of convergence sectors: 
the case of a cubic potential}

Consider the potential:
\be
\label{cubic}
W(z)=t_3z^3+t_2z^2~~.
\ee
This example appeared in the paper \cite{Klemm}, where it was used to carry
out a one-loop test of the Dijkgraaf-Vafa conjecture. As in \cite{Klemm}, we
assume for simplicity that $t_2$ and $t_3$ are real and positive, and write
them as $t_2=\frac{m}{2}$ and 
$t_3=\frac{g}{3}$ with positive $m$ and $g$. 
The potential has two local extrema along the real axis, namely a local 
minimum at $a_1=0$ and a local maximum at $a_2=-\frac{2t_2}{3t_3}=
-\frac{m}{g}<0$. 
Also note that $W(z)$ approaches $\pm \infty$ as $z$ approaches $\pm \infty$ 
along the real axis.

\subsection{Summary of the procedure of \cite{Klemm}}

The paper \cite{Klemm} follows \cite{DV1} by formulating a conjecture mapping
a one-matrix model based on the potential (\ref{cubic}) to the 
closed topological B-model on a certain non-compact Calabi-Yau space. When 
testing this, \cite{Klemm} performs a perturbative expansion
of this model around the two extrema $a_1$ and $a_2$, thereby
re-writing the theory as a a two-matrix model for
matrices $M_1,M_2$ with cubic 
potentials $W_1(M_1)$ and $W_2(M_2)$ and a 
logarithmic interaction term $W_{int}(M_1,M_2)$.
Since the point $a_2$ is a local {\em maximum} for the original potential $W$, 
this would produce a negative definite quadratic term in $W_2(M_2)$,
if one starts with a Hermitian matrix model,
thereby leading to an ill-defined perturbation expansion. To cure this
problem, \cite{Klemm} proposes that one should take $M_1$ to be
  anti-Hermitian and  $M_2$ to be Hermitian. With this redefinition,
the authors of \cite{Klemm} compute the first few perturbative terms 
and find agreement with the one-loop 
computation of the Kodaira-Spencer theory of the Calabi-Yau dual. 
We now show how this pragmatic procedure can be recovered 
in the holomorphic framework.

\subsection{Justification in terms of holomorphic matrix models}

Remember 
that a Hermitian  matrix model based on the cubic potential 
(\ref{cubic}) is ill-defined. 
This is due to the
exponential increase of the integrand for $\lambda_j\rightarrow -\infty$. 
On the other hand, the holomorphic matrix model leads to a meaningful
integral, provided that one chooses appropriate asymptotic sectors. 

Note that our definition 
(\ref{hmm}) uses the weight $\prod_{i\neq j}(\lambda_i-\lambda_j)$ 
rather than the weight $\prod_{i<j}(\lambda_i-\lambda_j)^2$ which is 
used in \cite{Klemm}. In this 
appendix, we shall temporarily switch to the representation:
\be
\label{evK}
{\cal Z}_N(\gamma,t)=
\int_\gamma d\lambda_1 \dots \int_\gamma d\lambda_N
\prod_{i<j}{(\lambda_i-\lambda_j)^2} e^{-N\sum_{j=1}^N{W(\lambda_j)}}~~
\ee
which differs from our convention (\ref{ev}) by a sign prefactor 
of $(-1)^{N(N-1)/2}$. This avoids some annoying sign 
factors when comparing with \cite{Klemm}.

\begin{figure}[hbtp]
\begin{center}
\scalebox{0.4}{\input{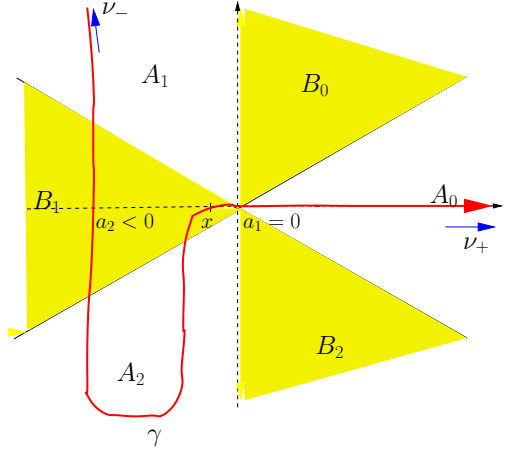}}
\end{center}
\caption{Convergence sectors for the case $deg~W=3$, 
$\theta_3=0$ and a good choice of contour belonging to the sector 
$(k_-,k_+)=(1,0)$.}
\label{3sectors}
\end{figure}

Thus we start with the partition function (\ref{evK}) for the potential
(\ref{cubic}). Since $t_3>0$, this model has $\theta_3=0$ and the asymptotic 
sectors shown in figure \ref{3sectors}. We claim that the correct partition 
function relevant for the work of \cite{Klemm} is 
${\cal Z}(1,0,t)$, associated with the asymptotic 
sectors $k_-=1$ and $k_+=0$. This is convergent by the
analysis of Section 2. 

In this set-up, the prescription of \cite{Klemm} can be recovered as follows 
(we shall neglect the gauge group volume prefactors, which are irrelevant 
here). 
Since the partition function (\ref{evK}) 
depends only on the asymptotic sectors of $\gamma$ 
(namely $\nu_-\in A_1$ and $\nu_+\in A_0$),
we are free to choose this curve as shown in figure \ref{3sectors}. 
This asymptotes to some line $d_-$ with tangent $\nu_-\in A_1$ 
for $t\rightarrow -\infty$, then 
follows a vertical line through the point $a_2=-m/g$, and finally
loops back to touch (and then follow) the real axis at some point $x$ lying in
between $a_2=-m/g$ and $a_1=0$. Thus an eigenvalue $\lambda$ 
sitting on $\gamma$ will 
be imaginary if it lies close to $a_2$ and real 
if it lies close
to $a_0$, thereby {\em naturally} 
implementing the requirement of \cite{Klemm}. The curve 
segments 
along $\gamma$ around the points $a_1$ and $a_2$ for which $\lambda$ has these
properties can be maximized by making the `well' at the bottom of
this curve to be very thin (i.e. take $x$ to be very close to $a_2$) and 
very deep.

Following \cite{Klemm}, we now expand the integral (\ref{evK}) (with $\gamma$
chosen as above) around the saddle point configurations:
\be
\lambda^{(0)}_i=a_1=0~~{\rm for}~~i=1\dots N_1~~,~~\lambda_j^{(0)}=a_2=-m/g
~~{\rm for}~~j=N_1+1\dots N~~
\ee
with $N_1$ some integer in the set $\{1\dots N\}$. 
Let $N_2:=N-N_1$. 
This gives:
\be
\label{Z_expansion}
{\cal Z}(t,\gamma)=\sum_{N_1+N_2=N}{\frac{N!}{N_1! N_2!}
\int_\gamma{d\mu_1}\dots \int_\gamma{d\mu_{N_1}}
\int_\gamma{d\nu_1}\dots \int_\gamma{d\nu_{N_2}}
\Delta(\mu,\nu) e^{-N S}} 
\ee
where we wrote 
$\lambda_i=\lambda_i^{(0)}+\mu_i$, $\lambda_{N_1+j}=\lambda_{N_1+j}^{(0)}+
\nu_j$
and:
\bea
\Delta(\mu,\nu)&=&
\prod_{1\leq i_1<i_2\leq N_1}{(\mu_{i_1}-\mu_{i_2})^2}
\prod_{1\leq j_1<j_2\leq N_1}{(\nu_{j_1}-\nu_{j_2})^2}
\prod_{i=1}^{N_1}\prod_{j=1}^{N_2}{(\mu_i-\nu_j+\frac{m}{g})^2}\nn\\
S&=&\sum_{i=1}^{N_1}{W(\mu_i)}-\sum_{j=1}^{N_2}{W(-\nu_j)}+
\frac{m^3}{6g^2}N_2~~.
\eea

Treating $\mu_i$ and $\nu_j$ as small fluctuations, we naturally 
have $\mu_i\in i\R$ and
$\nu_j\in \R$, since the eigenvalues $\lambda$ lying along $\gamma$ 
are imaginary in the vicinity of $a_2$ and real in the vicinity of $a_1$. 
Writing the last term in $\Delta$ as an exponential, one obtains a
logarithmic interaction potential\footnote{The authors of \cite{Klemm} 
further expand the logarithm as a power series in 
$\frac{g}{m}(\mu_i-\nu_j)$, a procedure which is justified if 
$|\frac{g}{m}(\mu_i-\nu_j)|<1$.}:
\be
W_{int}=2N_1N_2\ln\frac{m}{g}+2\sum_{i=1}^{N_1}\sum_{j=1}^{N_2}
\ln\left[1+\frac{g}{m}(\mu_i-\nu_j)\right]~~
\ee
and one-matrix potentials $W_1(\mu)=W(\mu)$ and $W_2(\nu)=-W(-\nu)$, which
allows one to write the result as an interacting two-matrix model \cite{Klemm}:
\be
Z=\sum_{N_1+N_2=N}{\frac{N!}{N_1!N_2!}
\int_{M_1^+=-M_1}{dM_1\int_{M_2^+=+M_2}{dM_2
e^{-N tr\left[W_1(M_1)+W_2(M_2)\right]+W_{int}(M_1,M_2)}}}}~~,
\ee
where the first integral is over anti-Hermitian matrices while the second
integral is over Hermitian matrices. Note that anti-Hermiticity of $M_1$ 
arises {\em naturally} in the holomorphic set-up.
This gives a conceptual justification for 
the procedure of \cite{Klemm}. Note also that we have provided a 
non-perturbative construction of the model involved in that work: 
it is simply the holomorphic matrix model with potential (\ref{cubic}), 
considered in the `phase' $(k_-,k_+)=(1,0)$.

\section{Derivation of the planar constraints for the $A_2$ model}

Let us show how the 
non-hyperelliptic Riemann surface expected from  the observations of 
\cite{DV3} arises 
in the holomorphic $A_2$ model. For this, we derive two algebraic 
constraints which hold in the planar limit, as a consequence of the planar 
equations of motion (\ref{A2eom0}). 

To derive the equations of interest, we shall use the partial
fraction decompositions:
\be
\label{pf1}
\frac{1}{(z-u)(z-v)}=\frac{1}{u-v}\left[\frac{1}{z-u}-\frac{1}{z-v}\right]
\ee
and:
\be
\label{pf2}
\frac{1}{(z-u)(z-v)(z-w)}=
\frac{1}{(u-v)(u-w)}\frac{1}{z-u}+
\frac{1}{(v-u)(v-w)}\frac{1}{z-v}+
\frac{1}{(w-u)(w-v)}\frac{1}{z-w}~~.
\ee

\subsection{The first constraint}

Using (\ref{pf1}), one  finds:
\bea
\label{omega_squared}
\omega_0^{(1)}(z)^2&=&2\int{ds}\int{ds'\frac{\rho_0^{(1)}(s)\rho_0^{(1)}(s')}{
\lambda(s)-\lambda(s')}\frac{1}{z-\lambda(s)}}~~\nn\\
\omega_0^{(2)}(z)^2&=&2\int{ds}\int{ds'\frac{\rho_0^{(2)}(s)\rho_0^{(2)}(s')}{
\lambda(s)-\lambda(s')}\frac{1}{z-\lambda(s)}}~~\\
\omega_0^{(1)}(z)\omega_0^{(2)}(z)&=&
\int{ds}\int{ds'\left[
\frac{\rho_0^{(1)}(s)\rho_0^{(2)}(s')}{\lambda(s)-\lambda(s')}
\frac{1}{z-\lambda(s)}+\frac{\rho_0^{(2)}(s)\rho_0^{(1)}(s')}{\lambda(s)-\lambda(s')}
\frac{1}{z-\lambda(s)}\right]}~~\nn
\eea
Combining these equations gives:
\be
\label{rel1intmd}
\omega_0^{(1)}(z)^2-\omega_0^{(1)}(z)\omega_0^{(2)}(z)+\omega_0^{(2)}(z)^2=
\int{ds \rho_0^{(1)}(s)\frac{W_1'(\lambda(s))}{z-\lambda(s)}}+
\int{ds \rho_0^{(2)}(s)\frac{W_2'(\lambda(s))}{z-\lambda(s)}}~~.
\ee
To arrive at this relation, we decomposed the integrals over $ds$ 
in (\ref{omega_squared}) into the pieces corresponding  
to the cuts $C_{13}^a$, 
$C_{23}^b$ and $C_{12}^c$. Then equation (\ref{rel1intmd}) results 
upon combining these 
pieces appropriately and performing the $s'$ integral by using the 
planar equations of motion (\ref{A2eom0}) and relations (\ref{rho_sym}). 
We next write (\ref{rel1intmd}) in the form:
\be
\label{first}
\omega_0^{(1)}(z)^2-\omega_0^{(1)}(z)\omega_0^{(2)}(z)+\omega_0^{(2)}(z)^2-W_1'(z)\omega_0^{(1)}(z)-
W_2'(z)\omega_0^{(2)}(z)+f^{(1)}_0(z)+f^{(2)}_0(z)=0~~,
\ee
where we used the planar equations of motion (\ref{A2eom0}) and the
definition of $\omega_0^{(\alpha)}(z)$ and we introduced 
the polynomials:
\bea
\label{foo}
f_0^{(1)}(z)&:=&\int{ds
  \rho_0^{(1)}(s)\frac{W_1'(z)-W_1'(\lambda(s))}{z-\lambda(s)}}\nn\\
f_0^{(2)}(z)&:=&\int{ds
  \rho_0^{(2)}(s)\frac{W_2'(z)-W_2'(\lambda(s))}{z-\lambda(s)}}~~.
\eea

\subsection{The second constraint}

To derive the second constraint, we use (\ref{pf2}) to compute:
\bea
\omega_0^{(1)}(z)^2\omega_0^{(2)}(z)&=&\int{d\alpha}\int{d\beta}\int{d\gamma 
\frac{2\rho_0^{(1)}(\alpha)\rho_0^{(1)}(\beta)\rho_0^{(2)}(\gamma)}{
(\lambda(\alpha)-\lambda(\beta))(\lambda(\alpha)-\lambda(\gamma))
(z-\lambda(\alpha))}}\nn\\
&+&\int{d\alpha}\int{d\beta}\int{d\gamma\frac{\rho_0^{(2)}(\alpha)\rho_0^{(1)}(\beta)
\rho_0^{(1)}(\gamma)}{(\lambda(\alpha)-\lambda(\beta))(\lambda(\alpha)-
\lambda(\gamma))(z-\lambda(\alpha))}}~~,
\eea
where we used redefinitions of $(\alpha,\beta, \gamma)$ by permutations 
to bring the right hand side to a convenient form.
Combining this with the equation obtained by permuting 
the indices $1\leftrightarrow 2$ gives:
\be
\omega_0^{(1)}(z)^2\omega_0^{(2)}(z)-
\omega_0^{(1)}(z)\omega_0^{(2)}(z)^2=
\int{d\alpha}\int{d\beta
W_1'(\lambda(\alpha))\frac{\rho_0^{(1)}(\alpha)\rho_0^{(2)}(\beta)}{(
\lambda(\alpha)-\lambda(\beta))(z-\lambda(\alpha))}}
-(1\leftrightarrow 2)~~,
\ee
where we used the planar equations of motion (\ref{A2eom0}) to perform the
integral over $\beta$. Defining the polynomials:
\bea
g^{(1)}_0(z)&:=&
\int{d\alpha}\int{d\beta\left[W'_1(z)-W'_1(\lambda(\alpha))\right]
\frac{\rho_0^{(1)}(\alpha)\rho_0^{(2)}(\beta)}{(\lambda(\alpha)-\lambda(\beta))
(z-\lambda(\alpha))}}~~,\nn\\
g^{(2)}_0(z)&:=&
\int{d\alpha}\int{d\beta\left[W'_2(z)-W'_2(\lambda(\alpha))\right]
\frac{\rho_0^{(2)}(\alpha)\rho_0^{(1)}(\beta)}{(\lambda(\alpha)-\lambda(\beta))
(z-\lambda(\alpha))}}~~,
\eea
we find:
\be
\label{intmd}
\omega_0^{(1)}(z)^2\omega_0^{(2)}(z)-
\omega_0^{(1)}(z)\omega_0^{(2)}(z)^2+g^{(1)}_0(z)-g^{(2)}_0(z)-
W_1'(z)U_1(z)+W_2'(z)U_2(z)=0~~,
\ee
where:
\bea
U_1(z)&:=&\int{d\alpha}\int{d\beta\frac{\rho_0^{(1)}(\alpha)\rho_0^{(2)}(\beta)}{
(\lambda(\alpha)-\lambda(\beta))(z-\lambda(\alpha))}}\nn\\
U_2(z)&:=&\int{d\alpha}\int{d\beta\frac{\rho_0^{(2)}(\alpha)\rho_0^{(1)}(\beta)}{
(\lambda(\alpha)-\lambda(\beta))(z-\lambda(\alpha))}}~~.
\eea
Using the equations of motion (\ref{A2eom0}), these quantities can be written:
\bea
U_1(z)&=&f_0^{(1)}(z)-W'_1(z)\omega_0^{(1)}(z)+\omega_0^{(1)}(z)^2~~\nn\\
U_2(z)&=&f_0^{(2)}(z)-W'_2(z)\omega_0^{(2)}(z)+\omega_0^{(2)}(z)^2~~.
\eea
Therefore, equation (\ref{intmd}) becomes:
\be
\label{second}
\omega_0^{(1)}(z)^2\omega_0^{(2)}(z)-
W_1'(z)[\omega_0^{(1)}(z)^2-W_1'(z)\omega_0^{(1)}(z)
+f^{(1)}_0(z)]+g^{(1)}_0(z)-
(1\leftrightarrow 2)=0~~.
\ee


\begin{thebibliography}{100}
\bibitem{DV1}{ R.~Dijkgraaf, C.~Vafa, 
{\em Matrix Models, Topological Strings, and Supersymmetric Gauge Theories},
 hep-th/0206255, Nucl. Phys. {\bf B644} (2002) 3-20.}
\bibitem{D_ft}{R.~Dijkgraaf, M.~T.~Grisaru, C.~S.~Lam, C.~Vafa, D.~Zanon, 
{\em Perturbative Computation of Glueball Superpotentials}, 
hep-th/0211017.}
\bibitem{DV3}{R.~Dijkgraaf, C.~Vafa, 
{\em On Geometry and Matrix Models},  Nucl.Phys. B644 (2002) 21-39, 
hep-th/0207106.}
\bibitem{Witten}{F.~Cachazo, M.~R.~Douglas, N.~Seiberg, E.~Witten,
{\em Chiral Rings and Anomalies in Supersymmetric Gauge Theory}, 
hep-th/0211170, JHEP {\bf 0212} (2002) 071.}
\bibitem{Seiberg}{N.~Seiberg, 
{\em  Adding Fundamental Matter to ``Chiral Rings and Anomalies in
  Supersymmetric Gauge Theory''}, hep-th/0212225.}
\bibitem{Klemm}{ A.~Klemm, M.~Marino, S.~Theisen, 
{\em Gravitational corrections in supersymmetric gauge theory and matrix 
models}, hep-th/0211216.}
\bibitem{Witten_CS}{E.~Witten, {\em Chern-Simons gauge theory as a string
theory}, The Floer memorial volume, 637--678, Progr. Math., 133, Birkhauser,
Basel, 1995, hep-th/9207094.}
\bibitem{Corrado}{S.~K.~Ashok, R.~Corrado, N.~Halmagyi, 
K.~D.~Kennaway, C.~Romelsberger, {\em Unoriented Strings, Loop Equations, and
  N=1 Superpotentials from Matrix Models}, hep-th/0211291.}
\bibitem{Whitham}{L.Chekhov, A.Mironov, 
{\em Matrix models vs. Seiberg-Witten/Whitham theories}, 
hep-th/0209085, Phys.Lett. {\bf B552} (2003) 293-302.}
\bibitem{Kostov}{I.~K. Kostov, 
{\em Conformal Field Theory Techniques in Random Matrix models}, 
hep-th/9907060.}
\bibitem{DSM}{R.~Dijkgraaf, A.~Sinkovics, M.~Temurhan,
{\em Matrix Models and Gravitational Corrections}, hep-th/0211241.}
\bibitem{Kontsevich}{M.~Kontsevich, {\em Intersection theory 
on the moduli space of curves and the matrix Airy function},
Commun. Math. Phys, {\bf 147}(1992)1.}
\bibitem{Witten2}{F.~Cachazo, N.~Seiberg, E.~Witten, 
{\em Phases of N=1 Supersymmetric Gauge Theories and Matrices}, 
hep-th/0301006.}
\bibitem{DV2}{R.~Dijkgraaf, C.~Vafa, 
{\em A Perturbative Window into Non-Perturbative Physics}, hep-th/0208048.}
\bibitem{Klemm2}{M.~Aganagic, A.~Klemm, M.~Marino, C.~Vafa,
{\em Matrix Model as a Mirror of Chern-Simons Theory}, hep-th/0211098.} 
\bibitem{Gorski}{A.~Gorsky, {\em Konishi anomaly and N=1 effective 
superpotentials from matrix models}, hep-th/0210281.}
\bibitem{Whitham2}{L.~Chekhov, A.~Marshakov, A.~Mironov, D.~Vasiliev,
{\em DV and WDVV}, hep-th/0301071.}
\bibitem{Whitham3}{A. Mironov, {\em N=1 SUSY inspired Whitham prepotentials 
and WDVV}, hep-th/0301196.}
\bibitem{complex}{T.~R.~Morris, 
{\em Checkered surfaces and complex matrices}, Nucl. Phys. {\bf B356}
(1991)703.}
\bibitem{Jurk}{J.~Jurkiewicz, 
{\em Regularization of one-matrix models}, 
Phys. Lett {\bf B245}(1990)178.} 
\bibitem{BCOV}{ M.~Bershadsky, S.~Cecotti, H.~Ooguri, C.~Vafa, 
{\em Kodaira-Spencer Theory of Gravity and Exact Results for Quantum String 
Amplitudes}, Commun. Math. Phys. {\bf 165} (1994) 311-428,hep-th/9309140.}
\bibitem{Kostov_ADE}{I. K. Kostov, 
{\em Solvable statistical models on a random lattice}, 
Nucl.Phys.Proc.Suppl. {\bf 45A} (1996) 13-28, hep-th/9509124;
{\em Gauge Invariant Matrix Model for the \^A-\^D-\^E Closed Strings}, 
Phys.Lett. {\bf B297} (1992) 74-81, hep-th/9208053.}
\bibitem{Cachazo}{ F. Cachazo, B. Fiol, K. Intriligator, S. Katz, C. Vafa, 
{\em A geometric unification of dualities}, Nucl.Phys. {\bf B628} 
(2002) 3-78, hep-th/0110028; 
 F. Cachazo, S. Katz, C. Vafa, {\em 
Geometric Transitions and N=1 Quiver Theories}, hep-th/0108120;
 F. Cachazo, K. Intriligator, C. Vafa, {\em 
A Large N Duality via a Geometric Transition}, Nucl.Phys. 
{\bf B603} (2001) 3-41, hep-th/0103067.}
\bibitem{Radu}{K.~Oh and R.~Tatar,
{\em Duality and confinement in N = 1 supersymmetric theories from 
geometric  transitions}, Adv.\ Theor.\ Math.\ Phys.\  {\bf 6}, 141 (2003), 
hep-th/0112040.}
\bibitem{DV4}{R.~Dijkgraaf, C.~Vafa, {\em N=1 Supersymmetry, Deconstruction, 
and Bosonic Gauge Theories}, hep-th/0302011.}
\bibitem{DV5}{ R.~Dijkgraaf, A.~Neitzke, C.~Vafa, 
{\em Large N Strong Coupling Dynamics in Non-Supersymmetric Orbifold Field 
Theories},  hep-th/0211194.}
\bibitem{Tatar}{R.~Roiban, R.~Tatar, J.~Walcher, 
{\em  Massless Flavor in Geometry and Matrix Models}, hep-th/0301217.}
\bibitem{us}{A.~Klemm, K.~Landsteiner, C.~I.~Lazaroiu and I.~Runkel,
{\em Constructing Gauge Theory Geometries from Matrix Models},
hep-th/0303032.} 
\end{thebibliography}
\end{document}